\documentclass[12pt]{article}

\usepackage{amsmath}
\usepackage[dvips]{graphicx}

\newcommand{\N}{N\raise.7ex\hbox{\underline{$\circ $}}$\;$}
\begin{document}

\title{On relativistic theory of spinning and deformable particles}
\author{A.N. Tarakanov
\thanks{E-mail: tarak-ph@mail.ru }\\
{\small Minsk State High Radiotechnical College}\\
{\small Independence Avenue 62, 220005, Minsk, Belarus} }
\date{}
\maketitle
\begin{abstract}

A model of relativistic extended particle is considered with the
help of generalization of space-time interval. Ten additional
dimensions are connected with six rotational and four
deformational degrees of freedom. An obtained 14-dimensional space
is assumed to be an embedding one both for usual space-time and
for 10-dimensional internal space of rotational and deformational
variables. To describe such an internal space relativistic
generalizations of inertia and deformation tensors are given.
Independence of internal and external motions from each other
gives rise to splitting the equation of motion and some conditions
for 14-dimensional metric. Using the 14-dimensional ideology makes
possible to assign a unique proper time for all points of extended
object, if the metric will be degenerate. Properties of an
internal space are discussed in details in the case of absence of
spatial rotations.
\end{abstract}

\section{ Introduction}

\parindent=24pt\ \; \; \; More and more attention is spared
to relativistic description of extended objects, which could serve
a basis for construction of dynamics of interacting particles.
Necessity of introduction extended objects to elementary particle
theory is out of doubt. Therefore, since H.A.Lorentz attempts to
introduce particles of finite size were undertaken. However a
relativization of extended body is found prove to be a difficult
problem as at once there was a contradiction to Einstein's
relativity principle. Even for simplest model of absolutely rigid
body~\cite{Dis} it is impossible for all points of a body to
attribute the same proper time. Therefore considering of
elementary particles with rotational degrees of freedom is often
incorrect, when it is supposed that the particle possesses
infinitesimal spatial sizes for this leads to formal treating of
internal degrees of freedom~\cite{Fre}-\cite{Tak3}. Rotational
degrees of freedom are degenerated in general, and as physical
variables remain only internal angular
momenta~\cite{Wey},~\cite{Dix1}. This difficulty has not been
overcame neither in extended electron theory by Markov~\cite{Mar}
nor in bilocal and multilocal theories by Yukawa~\cite{Yuk} and
Takabayashi~\cite{Tak4} nor in relativistic rotator by
Nakano~\cite{Nak}, causing large flow of works, nor in many other
works devoted to extended objects~\cite{Hara} and relativistic
continuous media~\cite{Grot}. It does not affect the further
development of ideas, but one cannot say anything about the sizes
of particles. At large distances from particle the account only
internal momenta can appear sufficient, but considering of
interactions must be correct only with taking into account
particle sizes.

\parindent=24pt\ Usually the description of extended particle is
carried out by means of consideration the moving 4-hedron,
$e^{\mu} = \{e^{\mu}_{(\lambda)}\}$, connected with a point inside
of object~\cite{Nak}. Thus, angular velocity $\Omega^{[\mu\nu]}$
is defined according to the formula~\cite{Tul1}
$$
  \Omega^{[\mu\nu]} = \eta^{(\lambda)(\kappa)} e^{\mu}_{(\lambda)}
  \frac{{D_{\mathrm{g}}} e^{\nu}_{(\kappa)}}{c d \tau} =
  \eta^{(\lambda)(\kappa)} e^{\mu}_{(\lambda)} \left [ \frac{d
  e^{\nu}_{(\kappa)}}{cd\tau} + \Gamma^{nu}_{\rho\sigma}
  \dot{y}^{\sigma} e^{\rho}_{(\kappa)} \right],\eqno{(1.1)}
$$
where $y^{\mu}$ are coordinates of a point with which the
reference point is connected, $\dot{y}^{\mu} = dy^{\mu} / cd \tau$
,
$$
  \Gamma^{\nu}_{\rho\sigma} = \frac{1}{2} g^{\nu\tau}
  [\partial_{\rho} g_{\tau\sigma} + \partial_{\sigma} g_{\rho\tau}
  - \partial_{\tau} g_{\rho\sigma}] \; ,\eqno{(1.2)}
$$
$\partial_{\rho} = \partial / \partial y^{\rho}$, $g_{\mu\nu}$ is
the metrics describing complex movement of this point.

\parindent=24pt\ Other approach is the description of extended
bodies in General Relativity by means of specifying
energy-momentum tensor and its moments~\cite{Tul1}-\cite{Ryab},
going back to Mathisson~\cite{Math} and Papapetrou~\cite{Pap}.
However here even in the elementary case of a free rotating
particle in flat space-time its momentum and velocity appear to be
not parallel, what is not quite clear in case of the isolated
particle. With the certain kind of Lagrangian first approach
proves to be equivalent to the second one~\cite{Tul1}.

\parindent=24pt\ In this work another attempt is done to describe
in details a relativistic extended particle by means of
introducing internal rotational and deformational degrees of
freedom. Our approach is based on dimensional extension of
space-time interval connected with a world line of any point
inside the particle. In this connection it is necessary to note
the work~\cite{Kam} in which multidimensional space is used for
the description of extended objects, and as an example the
relativistic string is considered. In our work we connect
additional measurements with both six rotation and hyper-rotation
angles and four deformational degrees of freedom as it was
partially made in~\cite{Izma}.

\parindent=24pt\ In our opinion, many works, devoted to extended
elementary particles
(e.g.,~\cite{Izma}-\cite{Tak3},~\cite{Mar}-\cite{Hara}, etc.),
spare a little attention to classical description. Partly it is
caused unobservability internal movements and connected with
traditional attributing to microobjects of quantum properties.
Passage to quantization is represented by natural way, but
investigation of classical opportunities should be sufficiently
complete and comprehensive. In this sense the description of
extended bodies from the general relativistic
viewpoint~\cite{Dix1} is represented more consecutive. Here we
also give only classical consideration. In Section 2 a
generalization of space-time interval, connected with any point
inside of an extended particle, is specified with the help of
introducing internal rotational and deformational degrees of
freedom. For such a particle we also define a notion of the
"center of inertia" as a point representing the motion of the
particle considered as a whole object. Relativistic inertia and
deformation tensors, as well as general equations of motion are
considered in Section 3. In Section 4 we split equations of motion
and obtain some conditions for metrics with the help of embedding
space formalism. In particular, certain condition imposed on
angular and deformational velocities gives rise to possibility of
introducing the unique proper time in the whole internal region of
extended object. Section 5 shows that equations of motion obtained
in Section 4 may be derived also under consideration of
degenerated 14-dimensional metric, whose rank equals to the rank
of the background 4-dimensional space-time. Properties of internal
space are discussed and its metric is derived in Section 6 in the
case of absence of spatial rotations ($\Omega^{[mn]} = 0$). Here
we also clarify the physical meaning of components $\Omega^{[0n]}$
of 4-dimensional angular velocity $\Omega^{[\mu\nu]}$, which are
proportional to corresponding components of angular momentum of
the internal point relative to the "center of inertia". This fact
allows to connect $\Omega^{[0n]}$ with spin of the particle
determined in Section 5 and redetermined in Section 6. Section 7
contains conclusive notes.

\section{ Rotating and deformable medium}

\parindent=24pt\ \; \; \; Let there is some space-time
(finite or infinite) domain, filled by some substance. We shall
assume this domain to be moving in space-time and therefore it is
possible to choose some point C, said to be the "center of
inertia", which represents movement of the domain as a whole.
Because the concept of the center of inertia cannot be uniquely
defined in a relativistic case~\cite{Pry} we shall use it
conditionally, implying as the "center of inertia" certain
preferred point inside domain, whose movement in space-time
defines movement of the domain as a whole. Domain, defined as a
part of the Universe, may be both any physical system and
fundamental particle. Here we consider such domain as a particle.

\parindent=24pt\ Let the observer be placed in the point O
(origin of coordinates) of space-time. Then coordinates of the
"center of inertia" C relative to O we shall denote through
$x^{\mu}$, coordinates of a point M inside particle relative to O
-- through $y^{\mu}$, and coordinates of the point M relative to
the "center of inertia" C -- through $r^{\mu}$.

\parindent=24pt\ We wish to consider a rotating particle, inside
of which internal substance undergoes some movement, so that a
field of 4-velocities is specified inside of domain occupied by
particle. Intuitively clearly, that world line of any point M
should wind around a world line of the point C. If one postulate,
that any movement always can be presented as geodesic movement
(see, e.g.,~\cite{DeSit1}-\cite{Pet}), it means, that spaces,
associated with movement of the "center of inertia" C and with
movement of point M, are various and possess various metrics. The
metric of the space, associated with point M, is complicated even
when the background space and space of substance are Minkowski
spaces. Should rotations and deformations be absent, these spaces
would coincide, and it would possible to write
$y^{\mu}=x^{\mu}+r^{\mu}$. In general case $y^{\mu}$ must be
considered as some functions from $x^{\mu}$, $r^{\mu}$ and
rotation angles $\varphi^{\mu\nu}$. In the further we shall
distinguish the indices connected with various variables, and
instead of $x^{\mu}$, $r^{\alpha}$, $\varphi^{\mu\nu}$ we shall
write $x^{(\mu)}$, $r^{\Hat{\alpha}}$, $\varphi^{[\mu\nu]}$. Thus,
indices in parentheses, $(\mu)$, $(\nu)$, $(\lambda)$, ..., are
concerned to the "center of inertia" C, moving in background
space, indices with hats, $\Hat{\alpha}$, $\Hat{\beta}$,
$\Hat{\gamma}$, ..., are concerned to any point M of the
substance, moving under internal deformations, while indices
without any brackets, $\mu$, $\nu$, $\lambda$, ..., are concerned
to any point M of the substance, moving relative to background
space by trajectory being geodesic line in the internal space of
the particle.

\parindent=24pt\ Infinitesimal variation of coordinates $y^{\mu}$
of internal point M, is obviously defined by expression
$$
  dy^{\mu} = \partial_{(\nu)}y^{\mu}dx^{(\nu)} +
  \partial_{\Hat{\alpha}}y^{\mu}dr^{\Hat{\alpha}}
  + \frac{1}{2} \frac{\partial y^{\mu}}{\partial
  \varphi^{[\lambda\kappa]}} d\varphi^{[\lambda\kappa]} \;
  .\eqno{(2.1)}
$$
On the other hand, $dy^{\mu}$ is caused by infinitesimal changes
$Dx^{(\mu)}$ and $Dr^{\Hat{\alpha}}$ of coordinates $x^{(\mu)}$
and $r^{\Hat{\alpha}}$, taking into account rotations and
deformation of substance. $Dx^{(\mu)}$ is caused, firstly, by
movement of the particle as a whole, and, secondly, by moving of
substance within the vicinity of the point C. $Dr^{\Hat{\alpha}}$
does not depend on external movement and is caused only by
rotation of particle and moving of substance within the vicinity
of the point M. Therefore it is possible to write
$$
  Dx^{(\mu)} = dx^{(\mu)} + V^{(\mu)}_{\mathrm{C}} cd\tau \;
  ,\eqno{(2.2)}
$$
$$
  Dr^{\Hat{\alpha}} = \frac{1}{2} (\delta^{\Hat{\alpha}}_{\cdot \;
  (\lambda)} \delta_{\Hat{\beta} (\kappa)} -
  \delta^{\Hat{\alpha}}_{\cdot \; (\kappa)} \delta_{\Hat{\beta}
  (\lambda)}) r^{\Hat{\beta}} d\varphi^{[\kappa\lambda]} +
  V^{\Hat{\alpha}}_{\mathrm{M}} cd\tau \; ,\eqno{(2.3)}
$$
where
$$
  V^{(\mu)}_{\mathrm{C}} =
  \delta^{(\mu)}_{\Hat{\alpha}} \left.
  \frac{dr^{\Hat{\alpha}}}{cd\tau}\;
  \right |_{r^{\Hat{\alpha}} = 0 , \; d\varphi^{[\kappa\lambda]} =
  0}\eqno{(2.4)}
$$
is a velocity of substance within the vicinity of the "center of
inertia" C due to a deformation of substance,
$$
  V^{\Hat{\alpha}}_{\mathrm{M}} = \left.
  \frac{dr^{\Hat{\alpha}}}{cd\tau} \;
  \right |_{d\varphi^{[\kappa\lambda]} = 0}\eqno{(2.5)}
$$
is a velocity of substance within the vicinity of the point M due
to a deformation of substance, $\tau$ is proper time of the point
M, $\delta^{(\mu)}_{\Hat{\alpha}}$ are usual Kronecker symbols
($\delta^{(\mu)}_{\Hat{\alpha}}=1$ when $\alpha=\mu$ and
$\delta^{(\mu)}_{\Hat{\alpha}}=0$ when $\alpha \neq \mu$). Thus,
movement of the point M is characterized by infinitesimal variable
$$
  dy^{\mu} = \delta^{\mu}_{\cdot \; (\lambda)} Dx^{(\lambda)} +
  \delta^{\mu}_{\cdot \; \Hat{\alpha}} Dr^{\Hat{\alpha}} =
  \delta^{\mu}_{\cdot \; (\lambda)} dx^{(\lambda)} +
\nonumber
$$
$$
  + \frac{1}{2} (\delta^{\mu}_{\cdot \; (\lambda)}
  \delta_{\Hat{\alpha}(\kappa)} - \delta^{\mu}_{\cdot \; (\kappa)}
  \delta_{\Hat{\alpha}(\lambda)}) r^{\Hat{\alpha}}
  d\varphi^{[\kappa\lambda]} + (\delta^{\mu}_{\cdot \; (\lambda)}
  V^{(\lambda)}_{\mathrm{C}} + \delta^{\mu}_{\cdot \;
  \Hat{\alpha}} V^{\Hat{\alpha}}_{\mathrm{M}}) \; cd\tau \;
  ,\eqno{(2.6)}
$$
where $\delta^{\mu}_{\cdot \; (\lambda)}$, $\delta^{\mu}_{\cdot \;
\Hat{\alpha}}$ are also Kronecker symbols. To eliminate unknown
quantities $V^{(\mu)}_{\mathrm{C}}$,
$V^{\Hat{\alpha}}_{\mathrm{M}}$, it is reasonable to assume, that
they are connected with each other by transformation of a kind
$$
  (\delta^{\mu}_{\cdot \; (\lambda)} V^{(\lambda)}_{\mathrm{C}} +
  \delta^{\mu}_{\cdot \; \Hat{\alpha}}
  V^{\Hat{\alpha}}_{\mathrm{M}}) \; cd\tau = f^{\mu}_{\cdot \;
  (\lambda)} dx^{\lambda} + b^{\mu}_{\cdot \; \Hat{\alpha}}
  dr^{\Hat{\alpha}} + \frac{1}{2} d^{\mu}_{\cdot \;
  [\lambda\kappa]} d\varphi^{[\lambda\kappa]} \; ,\eqno{(2.7)}
$$
where $f^{\mu}_{\cdot \; (\lambda)}$, $b^{\mu}_{\cdot \;
  \Hat{\alpha}}$ and $d^{\mu}_{\cdot \; [\lambda\kappa]}$ are some
  functions from coordinates $x^{(\mu)}$, covering external space,
  internal coordinates $r^{\Hat{\alpha}}$ and rotation angles
  $\varphi^{[\mu\nu]}$.

\parindent=24pt\ Comparing (2.1) and (2.6) and taking into account
(2.7), we have
$$
  \partial_{(\lambda)} y^{\mu} = \delta^{\mu}_{\cdot \; (\lambda)}
  + f^{\mu}_{\cdot \; (\lambda)} \doteq e^{\mu}_{\cdot \;
  (\lambda)} \; ,\eqno{(2.8)}
$$
$$
  \partial_{\Hat{\alpha}} y^{\mu} = b^{\mu}_{\cdot \;
  \Hat{\alpha}} \; ,\eqno{(2.9)}
$$
$$
  \frac{\partial y^{\mu}}{\partial \varphi^{[\lambda\kappa]}}
  \doteq d^{\mu}_{\cdot \; [\lambda\kappa]} - (\delta^{\mu}_{\cdot
  \; (\lambda)} \delta_{{\Hat{\alpha}(\kappa)}} -
  \delta^{\mu}_{\cdot \; (\kappa)}
  \delta_{{\Hat{\alpha}(\lambda)}}) r^{\Hat{\alpha}} \;
  .\eqno{(2.10)}
$$

\parindent=24pt\ For the description of movement of point M we
shall write down a corresponding squared interval that will look
like
$$
  dS^2 = \Sigma c^{2} d\tau^{2} = g_{\mu\nu} dy^{\mu} dy^{\nu} =
  g_{\mu\nu} e^{\mu}_{\cdot \; (\lambda)} e^{\nu}_{\cdot \;
  (\kappa)} dx^{(\lambda)} dx^{(\kappa)} + g_{\mu\nu}
  b^{\mu}_{\cdot \; \Hat{\alpha}} b^{\nu}_{\cdot \; \Hat{\beta}}
  dr^{\Hat{\alpha}} dr^{\Hat{\beta}} +
$$
$$
  + \frac{1}{4} g_{\mu\nu} \left [d^{\mu}_{\cdot \;
  [\lambda\kappa]} - (\delta^{\mu}_{\cdot \; (\lambda)}
  \eta_{(\kappa)\Hat{\alpha}} - \delta^{\mu}_{\cdot \; (\kappa)}
  \eta_{(\lambda)\Hat{\alpha}}) r^{\Hat{\alpha}} \right ] \cdot
$$
$$
  \cdot \left [d^{\nu}_{\cdot \; [\rho\sigma]} -
  (\delta^{\nu}_{\cdot \;
  (\rho)} \eta_{(\sigma)\Hat{\beta}} - \delta^{\nu}_{\cdot \;
  (\sigma)} \eta_{(\rho)\Hat{\beta}}) r^{\Hat{\beta}} \right ]
  d\varphi^{[\lambda\kappa]} d\varphi^{[\rho\sigma]} + 2
  g_{\mu\nu} e^{\mu}_{\cdot \; (\lambda)} b^{\nu}_{\cdot \;
  \Hat{\alpha}} dx^{(\lambda)} dr^{\Hat{\alpha}} +
$$
$$
  + g_{\mu\nu} e^{\mu}_{\cdot \; (\lambda)} \left [d^{\nu}_{\cdot \;
  [\rho\sigma]} - (\delta^{\nu}_{\cdot \; (\rho)}
  \eta_{(\sigma)\Hat{\beta}} - \delta^{\nu}_{\cdot \; (\sigma)}
  \eta_{(\rho)\Hat{\beta}}) r^{\Hat{\beta}} \right ]
  dx^{(\lambda)} d\varphi^{[\rho\sigma]} +
$$
$$
  + g_{\mu\nu} b^{\mu}_{\cdot \; \Hat{\alpha}} \left
  [d^{\nu}_{\cdot \; [\rho\sigma]} - (\delta^{\nu}_{\cdot \;
  (\rho)} \eta_{(\sigma)\Hat{\beta}} - \delta^{\nu}_{\cdot \;
  (\sigma)} \eta_{(\rho)\Hat{\beta}}) r^{\Hat{\beta}} \right ]
  dr^{\Hat{\alpha}} d\varphi^{[\rho\sigma]} \; ,\eqno{(2.11)}
$$
where $\Sigma = \pm 1$ or 0, and through $g_{\mu\nu}$ the metric
of space, associated with movement of point M is designated. Here
it should be noted that using the metric $g_{\mu\nu}$ is connected
with unknown in advance character of movement of point M. Should
the trajectory of point M be known, it would be possible to use
instead of $g_{\mu\nu}$ the metric of background space
$\eta_{(\mu)(\nu)} = \delta_{(\mu)}^{\; \cdot \; \lambda}
\delta_{(\nu)}^{\; \cdot \; \kappa} \eta_{\lambda\kappa}$ together
with a restriction in the form of the equation of a trajectory of
point M, which in case of using the metric $g_{\mu\nu}$ turns out
to be an integral of the equations of motion
$$
  \ddot{y}^{\mu} + \Gamma^{\mu}_{\nu\lambda} \dot{y}^{\nu}
  \dot{y}^{\lambda}= 0 \; .\eqno{(2.12)}
$$

\parindent=24pt\ Starting from expression (2.11) the "center
of inertia" may be defined as follows. Because movement of the
"center of inertia" should look as movement of a material point in
background space it is reasonable to put
$$
  g_{\mu\nu} e^{\mu}_{\; \cdot (\lambda)} e^{\nu}_{\; \cdot
  (\kappa)} = \eta_{(\lambda)(\kappa)} \; ,\eqno{(2.13)}
$$
$$
  \eta^{(\lambda)(\kappa)} e^{\mu}_{\; \cdot (\lambda)} e^{\nu}_{\; \cdot
  (\kappa)} = g^{\mu\nu}\eqno{(2.14)}
$$
for all values of $r^{\Hat{\alpha}}$, which are not leaving for
domain, occupied by substance. If $x^{(\mu)}$ are not coordinates
of the "center of inertia" then relations (2.13), (2.14) should
not be carried out.

\parindent=24pt\ Introduction of relativistic generalizations
of inertia and deformation tensors allows to write down the
interval (2.11) in more compact form. That will be a theme of the
following Section.

\section{ Inertia and deformation tensors}

\parindent=24pt\ \; \; \; Let us consider in more details
the quantities entering into the interval (2.11) before
differentials. First of all, for rotating particle we need to
enter a relativistic inertia tensor. We shall consider specific
quantities, i.e., quantities associated with a unit mass. It is
natural to do so, as for as the concept of mass and mass
distribution inside of considered substance still are not
determined, the more so in the inertia tensor for a material point
the mass of it is only constant multiplier with physical dimension
of mass.

\parindent=24pt\ To give a relativistic generalization of inertia
tensor, we shall start from nonrelativistic description of
absolutely rigid body. Proper inertia tensor of a unit mass
particle relative to the beginning of coordinate system in
three-dimensional space looks like
$$
  j_{ik} = {\bf r}^2 \delta_{ik} - r_{i} r_{k} \; ,\eqno{(3.1)}
$$
where $r_{k}$ are coordinates of the particle (see,
e.g.,~\cite{Lan}). Introduction of (specific) linear momentum of
the particle relative to the beginning of coordinate system,
$$
  i_{ik} = \varepsilon_{ikm} r_{m} \; ,\eqno{(3.2)}
$$
allows to write the inertia tensor (3.1) as product of the linear
momenta
$$
  j_{ik} = -j_{im} j_{mk} \; .\eqno{(3.3)}
$$

\parindent=24pt\ In a relativistic case for the description
of rotational movement except for three angles $\varphi^{12}$,
$\varphi^{23}$, $\varphi^{31}$, connected with rotations in
three-dimensional space, one should require three more angular
variables $\varphi^{0i}$, connected with hyper-rotations in planes
$(0i)$. Therefore, there should be also components of generalized
inertia tensor corresponding to these planes. As it is well known,
$\varepsilon_{ijk}$ is 0-component of four-dimensional tensor
density by Levy-Civita: $\varepsilon_{ijk} = \varepsilon_{0ijk}$.
Hence, instead of linear momentum tensor of the second rank (3.2)
in three-dimensional space there arises a tensor of the third rank
$$
  i^{0}_{\mu\nu\lambda} = \varepsilon_{\mu\nu\lambda\kappa}
  r^{\kappa}\eqno{(3.4)}
$$
in four-dimensional space, and, obviously, $i_{ik} = i^{0}_{0ik}$.
Accordingly, instead of inertia tensor of the second rank (3.3)
there arises a tensor of the fourth rank:
$$
  j^{0}_{\mu\nu , \lambda\kappa} = j^{0}_{\lambda\kappa , \mu\nu}
  = - \eta^{\alpha\beta} i^{0}_{\mu\nu\alpha}
  i^{0}_{\lambda\kappa\beta}= - \eta^{\alpha\beta}
  \varepsilon_{\mu\nu\alpha\rho}
  \varepsilon_{\lambda\kappa\beta\sigma}
  r^{\rho} r^{\sigma} = \eta_{\mu\nu\rho , \lambda\kappa\sigma}
  r^{\rho} r^{\sigma} =
$$
$$
  = [\eta_{\mu\lambda} \eta_{\nu\kappa}
  \eta_{\rho\sigma} + \eta_{\mu\kappa} \eta_{\nu\sigma}
  \eta_{\rho\lambda} + \eta_{\mu\sigma} \eta_{\nu\lambda}
  \eta_{\rho\kappa} - \eta_{\mu\kappa} \eta_{\nu\lambda}
  \eta_{\rho\sigma} - \eta_{\mu\lambda} \eta_{\nu\sigma}
  \eta_{\rho\kappa} - \eta_{\mu\sigma} \eta_{\nu\kappa}
  \eta_{\rho\lambda}] r^{\rho} r^{\sigma} =
$$
$$
  = [(\eta_{\mu\lambda}
  \eta_{\nu\kappa} - \eta_{\mu\kappa} \eta_{\nu\lambda})
  r^{2}_{\eta} + \eta_{\mu\kappa} r_{\nu} r_{\lambda} +
  \eta_{\nu\lambda} r_{\mu} r_{\kappa} - \eta_{\mu\lambda} r_{\nu}
  r_{\kappa} - \eta_{\nu\kappa} r_{\mu} r_{\lambda}] \; .\eqno{(3.5)}
$$
It should be noted, that this tensor can be written down in other
form:
$$
  j^{0}_{\mu\nu , \lambda\kappa} = - \frac{i}{2} C_{\mu\nu ,
  \lambda\kappa , \rho\sigma} D^{\rho\sigma} r^{2}_{\eta} \;
  ,\eqno{(3.6)}
$$
where
$$
  r^{2}_{\eta} = \eta_{\alpha\beta} r^{\alpha} r^{\beta} \;
  .\eqno{(3.7)}
$$
Quantities
$$
  C_{\mu\nu , \lambda\kappa , \rho\sigma} = i [\eta_{\mu\lambda}
  \eta_{\nu\rho} \eta_{\kappa\sigma} + \eta_{\mu\rho}
  \eta_{\nu\kappa} \eta_{\lambda\sigma} - \eta_{\mu\kappa}
  \eta_{\nu\rho} \eta_{\lambda\sigma} - \eta_{\mu\rho}
  \eta_{\nu\lambda} \eta_{\kappa\sigma}]\eqno{(3.8)}
$$
and
$$
  D^{\mu\nu} = \eta^{\mu\nu} - \frac{2 r^{\mu}
  r^{\nu}}{r^{2}_{\eta}}\eqno{(3.9)}
$$
satisfy to relations
$$
  \frac{1}{4} \eta^{\mu\nu , \lambda\kappa} C_{\mu\nu ,
  \lambda\kappa , \rho\sigma} = \frac{1}{4} ( \eta^{\mu\lambda}
  \eta{\nu\kappa} - \eta^{\nu\lambda} \eta{\mu\kappa}) C_{\mu\nu
  , \lambda\kappa , \rho\sigma} = i(d-1)\eta_{\rho\sigma} \;
  ;\eqno{(3.10)}
$$
$$
  \eta_{\rho\sigma} D^{\mu\rho} D^{\nu\sigma} = \eta_{\mu\nu} \;
  ,\eqno{(3.11)}
$$
$$
  D^{\mu\rho} D_{\rho\nu} = \delta^{\mu}_{\nu} \; ,\eqno{(3.12)}
$$
where
$$
  D_{\mu\nu} = \eta_{\mu\nu} - \frac{2 \eta_{\mu\rho}
  \eta_{\nu\sigma} r^{\rho} r^{\sigma}}{r^{2}_{\eta}} \; ,\eqno{(3.13)}
$$
$d = \eta^{\mu\nu} \eta_{\mu\nu} = D^{\mu\nu} D_{\mu\nu} = 4$ is
the dimension of the background space.

\parindent=24pt\ Let us note also, that if four-dimensional
background space is the Minkowski space ${\bf
E}^{\mathrm{R}}_{1,3}$, then quantities (3.8) are structural
constants of the Lorentz group. It is obvious, the relation that
connects three-dimensional inertia tensor with a four-dimensional
tensor is $j_{ik} = j^{0}_{0i , 0j}$.

\parindent=24pt\ A relativistic generalization of the inertia
tensor above is natural and differs from often used inertia tensor
of the second rank (see,
e.g.,~\cite{Izma},~\cite{Tul1},~\cite{Tul2}). There exists also an
attempt to express inertia tensor of the second rank through
tensor of the fourth rank~\cite{Pry}.

\parindent=24pt\ Taking into account various character of indices,
one should written (3.4) and (3.5) as
$$
  i^{0}_{[\mu\nu](\lambda)} =
  \varepsilon_{[\mu\nu][\lambda\kappa]} \delta^{(\kappa)}_{\;
  \cdot \Hat{\alpha}} r^{\Hat{\alpha}} \; ,\eqno{(3.14)}
$$
$$
  j^{0}_{[\mu\nu][\lambda\kappa]} = -\eta^{(\rho)(\sigma)}
  i^{0}_{[\mu\nu](\rho)} i^{0}_{[\lambda\kappa](\sigma)} =
  - \frac{i}{2} C_{[\mu\nu][\lambda\kappa](\rho)(\sigma)}
  D^{(\rho)(\sigma)} r^{2}_{\eta} \; ,\eqno{(3.15)}
$$
where
$$
  D^{(\rho)(\sigma)} = \eta^{(\rho)(\sigma)} - \frac{2
  \delta^{(\rho)}_{\; \cdot \Hat{\alpha}} \delta^{(\sigma)}_{\;
  \cdot \Hat{\beta}} r^{\Hat{\alpha}}
  r^{\Hat{\beta}}}{r^{2}_{\eta}} \; ,\eqno{(3.16)}
$$
$$
  r^{2}_{\eta} = \eta_{(\rho)(\sigma)} \delta^{(\rho)}_{\; \cdot
  \Hat{\alpha}} \delta^{(\sigma)}_{\; \cdot \Hat{\beta}}
  r^{\Hat{\alpha}} r^{\Hat{\beta}} = h^{0}_{\Hat{\alpha}
  \Hat{\beta}} r^{\Hat{\alpha}} r^{\Hat{\beta}} \; .\eqno{(3.17)}
$$

\parindent=24pt\ Let us define also dual quantities:
$$
  \Tilde{i}^{0}_{[\mu\nu](\lambda)} = \frac{1}{2}
  \varepsilon^{\; \cdot \; \cdot \; [\rho\sigma]}_{[\mu\nu]}
  i^{0}_{[\rho\sigma](\lambda)} = (\eta_{(\nu)(\lambda)}
  \eta_{(\mu)\Hat{\alpha}} - \eta_{(\mu)(\lambda)}
  \eta_{(\nu)\Hat{\alpha}}) r^{\Hat{\alpha}} \; ,\eqno{(3.18)}
$$
$$
  \Tilde{j}^{0}_{\widetilde{[\mu\nu]}[\lambda\kappa]} =
  \Tilde{j}^{0}_{[\lambda\kappa]\widetilde{[\mu\nu]}} =
  \frac{1}{2} \varepsilon^{\; \cdot \; \cdot \;
  [\rho\sigma]}_{[\mu\nu]} j^{0}_{[\rho\sigma][\lambda\kappa]} = -
  \eta^{(\rho)(\sigma)} \Tilde{i}^{0}_{[\mu\nu](\rho)}
  \Tilde{i}^{0}_{[\lambda\kappa](\sigma)} =
$$
$$
  =(i^{0}_{[\lambda\kappa](\mu)} \eta_{(\nu)\Hat{\alpha}} -
  i^{0}_{[\lambda\kappa](\nu)} \eta_{(\mu)\Hat{\alpha}})
  r^{\Hat{\alpha}} = (\varepsilon_{[\lambda\kappa][\mu\sigma]}
  \eta_{(\nu)\Hat{\alpha}} -
  \varepsilon_{[\lambda\kappa][\nu\sigma]}
  \eta_{(\mu)\Hat{\alpha}}) \delta^{(\sigma)}_{\; \cdot
  \Hat{\beta}} r^{\Hat{\alpha}} r^{\Hat{\beta}} \; ,\eqno{(3.19)}
$$
$$
  \Tilde{\Tilde{j}}^{0}_{[\mu\nu][\lambda\kappa]} = \frac{1}{4}
  \varepsilon^{\; \cdot \; \cdot \; [\rho\sigma]}_{[\mu\nu]}
  \varepsilon^{\; \cdot \; \cdot \;
  [\tau\omega]}_{[\lambda\kappa]} j^{0}_{[\rho\sigma][\tau\omega]}
  = \hspace{61mm}
$$
$$
  = -\eta^{(\rho)(\sigma)} \Tilde{i}^{0}_{[\mu\nu](\rho)}
  \Tilde{i}^{0}_{[\lambda\kappa](\sigma)} =
  -iC_{[\mu\nu][\lambda\kappa](\rho)(\sigma)} \delta^{(\rho)}_{\;
  \cdot \Hat{\alpha}} \delta^{(\sigma)}_{\; \cdot \Hat{\beta}}
  r^{\Hat{\alpha}} r^{\Hat{\beta}} = \hspace{1mm}
$$
$$
  = [\eta_{(\mu)\Hat{\alpha}} (\eta_{(\nu)(\kappa)}
  \eta_{(\lambda)\Hat{\beta}} - \eta_{(\nu)(\lambda)}
  \eta_{(\kappa)\Hat{\beta}}) - \hspace{33mm}
$$
$$
  - \eta_{(\nu)\Hat{\alpha}}
  (\eta_{(\mu)(\kappa)} \eta_{(\lambda)\Hat{\beta}} -
  \eta_{(\mu)(\lambda)} \eta_{(\kappa)\Hat{\beta}})]
  r^{\Hat{\alpha}} r^{\Hat{\beta}} \; .\hspace{20mm}\eqno{(3.20)}
$$

\parindent=24pt\ It is not difficult to show, that following
relations take place:
$$
  \frac{1}{4} \eta^{[\mu\nu][\lambda\kappa]}
  i^{0}_{[\mu\nu](\rho)} i^{0}_{[\lambda\kappa](\sigma)} =
  r^{2}_{\eta} N^{0}_{(\rho)(\sigma)} \; ,\eqno{(3.21)}
$$
$$
  \frac{1}{4} \eta^{[\mu\nu][\lambda\kappa]}
  \Tilde{i}^{0}_{[\mu\nu](\rho)}
  \Tilde{i}^{0}_{[\lambda\kappa](\sigma)} = r^{2}_{\eta}
  N^{0}_{(\rho)(\sigma)} \; ,\eqno{(3.22)}
$$
$$
  \frac{1}{4} \eta^{[\mu\nu][\lambda\kappa]}
  i^{0}_{[\mu\nu](\rho)} \Tilde{i}^{0}_{[\lambda\kappa](\sigma)} =
  0 \; ;\eqno{(3.23)}
$$
$$
  \frac{1}{4} \eta^{[\rho\sigma][\tau\omega]}
  j^{0}_{[\mu\nu][\rho\sigma]} j^{0}_{[\lambda\kappa][\tau\omega]}
  = r^{2}_{\eta} j^{0}_{[\mu\nu][\lambda\kappa]} \; ,\eqno{(3.24)}
$$
$$
  \frac{1}{4} \eta^{[\rho\sigma][\tau\omega]}
  \Tilde{\Tilde{j}}^{0}_{[\mu\nu][\rho\sigma]}
  \Tilde{\Tilde{j}}^{0}_{[\lambda\kappa][\tau\omega]} =
  -r^{2}_{\eta} \Tilde{\Tilde{j}}^{0}_{[\mu\nu][\lambda\kappa]} \;
  ,\eqno{(3.25)}
$$
$$
  \frac{1}{4} \eta^{[\rho\sigma][\tau\omega]}
  j^{0}_{[\mu\nu][\rho\sigma]}
  \Tilde{\Tilde{j}}^{0}_{[\lambda\kappa][\tau\omega]} = 0
  \; ,\eqno{(3.26)}
$$
where
$$
  \eta^{[\rho\sigma][\tau\omega]} = \eta^{(\rho)(\tau)}
  \eta^{(\sigma)(\omega)} - \eta^{(\rho)(\omega)}
  \eta^{(\sigma)(\tau)} \; ,\eqno{(3.27)}
$$
and quantity
$$
  N^{0}_{(\rho)(\sigma)} = \eta_{(\rho)(\sigma)} -
  \frac{\eta_{(\rho)\Hat{\alpha}} \eta_{(\sigma) \Hat{\beta}}
  r^{\Hat{\alpha}} r^{\Hat{\beta}}}{r^{2}_{\eta}} = \frac{1}{2}
  (D_{(\rho)(\sigma)} + \eta_{(\rho)(\sigma)})\eqno{(3.28)}
$$
satisfies to relation
$$
  \eta^{(\rho)(\sigma)} N^{0}_{(\mu)(\rho)} N^{0}_{(\sigma)(\nu)}
  = N^{0}_{(\mu)(\nu)} \; .\eqno{(3.29)}
$$

\parindent=24pt\ It is easy to see now, that quantity standing
before product of angular differentials in (2.11) is actually
represents the sum of twice dual inertia tensor
$\Tilde{\Tilde{j}}^{0}_{[\mu\nu][\lambda\kappa]}$ and some
additional terms. We shall introduce a denotation
$$
 \Tilde{\Tilde{j}}_{[\mu\nu][\lambda\kappa]} =
 \Tilde{\Tilde{j}}^{0}_{[\mu\nu][\lambda\kappa]} - g_{\rho\sigma}
 d^{\rho}_{\; \cdot [\mu\nu]} d^{\sigma}_{\; \cdot
 [\lambda\kappa]} - g_{\rho\sigma} \eta^{\rho(\tau)}
 \left[\Tilde{i}^{0}_{[\mu\nu](\tau)} d^{\sigma}_{\; \cdot
 [\lambda\kappa]} + \Tilde{i}^{0}_{[\lambda\kappa](\tau)}
 d^{\sigma}_{\; \cdot [\mu\nu]} \right] +
$$
$$
  + \left (g_{\rho\sigma} \eta^{\rho(\tau)} \eta^{\sigma(\omega)} -
  \eta^{(\tau)(\omega)} \right ) \Tilde{i}^{0}_{[\mu\nu](\tau)}
  \Tilde{i}^{0}_{[\lambda\kappa](\omega)} = \hspace{21mm}
$$
$$
  = -g_{\rho\sigma} \left [d^{\rho}_{\; \cdot [\mu\nu]} +
  \eta^{\rho(\tau)} \Tilde{i}^{0}_{[\mu\nu](\tau)} \right ]
  \left [d^{\sigma}_{\; \cdot [\lambda\kappa]} +
  \eta^{\sigma(\omega)} \Tilde{i}^{0}_{[\lambda\kappa](\omega)}
  \right ] \; . \hspace{13mm}\eqno{(3.30)}
$$
Then the quantity
$$
 j_{[\mu\nu][\lambda\kappa]} = \frac{1}{4}
  \varepsilon^{\; \cdot \; \cdot \; [\rho\sigma]}_{[\mu\nu]}
  \varepsilon^{\; \cdot \; \cdot \;
  [\tau\omega]}_{[\lambda\kappa]}
  \Tilde{\Tilde{j}}_{[\rho\sigma][\tau\omega]} = \hspace{80mm}
$$
$$
  = j^{0}_{[\mu\nu][\lambda\kappa]} -
 g_{\rho\sigma} \Tilde{d}^{\rho}_{\; \cdot [\mu\nu]}
 \Tilde{d}^{\sigma}_{\; \cdot [\lambda\kappa]} + g_{\rho\sigma}
 \eta^{\rho(\tau)} \left[i^{0}_{[\mu\nu](\tau)}
 \Tilde{d}^{\sigma}_{\; \cdot [\lambda\kappa]} +
 i^{0}_{[\lambda\kappa](\tau)} \Tilde{d}^{\sigma}_{\; \cdot
 [\mu\nu]} \right] +
$$
$$
  + \left (g_{\rho\sigma} \eta^{\rho(\tau)} \eta^{\sigma(\omega)} -
  \eta^{(\tau)(\omega)} \right ) i^{0}_{[\mu\nu](\tau)}
  i^{0}_{[\lambda\kappa](\omega)} = \hspace{36mm}
$$
$$
  = -g_{\rho\sigma} \left [\eta^{\rho(\tau)}
  i^{0}_{[\mu\nu](\tau)} - \Tilde{d}^{\rho}_{\; \cdot [\mu\nu]}
  \right ] \left [\eta^{\sigma(\omega)}
  i^{0}_{[\lambda\kappa](\omega)} - \Tilde{d}^{\sigma}_{\; \cdot
  [\lambda\kappa]} \right ] \hspace{28mm}\eqno{(3.31)}
$$
will be called a {\it dynamical inertia tensor} of the point M
relative to the "center of inertia" C.

\parindent=24pt\ This definition differs from usual one
({\it a geometrical inertia tensor}) (3.1), (3.5) because the
quantity (3.23) is defined by not only geometrical characteristics
(coordinates $r^{\Hat{\alpha}}$) of point M, but also physical
processes in a vicinity of this point. This is due to presence of
quantities $d^{\mu}_{\; \cdot [\lambda\kappa]}$ in (3.23). If some
closed domain bound the substance under consideration, then
inertia tensor of the whole such an object can be obtained with
multiplication of (3.23) by mass density and subsequent
integration by volume of the domain.

\parindent=24pt\ Due to relations (2.13), (2.14) dynamical inertia
tensor $j_{[\mu\nu][\lambda\kappa]}$ and dual tensor
$\Tilde{\Tilde{j}}_{[\mu\nu][\lambda\kappa]}$  can be written down
in the form
$$
   j_{[\mu\nu][\lambda\kappa]} = -\eta^{(\rho)(\sigma)}
   i_{[\mu\nu](\rho)} i_{[\lambda\kappa](\sigma)} \; ,\eqno{(3.32)}
$$
$$
   \Tilde{\Tilde{j}}_{[\mu\nu][\lambda\kappa]} =
   -\eta^{(\rho)(\sigma)} \Tilde{i}_{[\mu\nu](\rho)}
   \Tilde{i}_{[\lambda\kappa](\sigma)} \; ,\eqno{(3.33)}
$$
Quantity
$$
  i_{[\mu\nu](\lambda)} = \eta_{(\lambda)(\kappa)}
  e^{(\kappa)}_{\; \cdot \rho} \left [ \eta^{\rho(\tau)}
  i^{0}_{[\mu\nu](\tau)} - \Tilde{d}^{\rho}_{\; \cdot [\mu\nu]}
  \right ]\eqno{(3.34)}
$$
will be called {\it a dynamical momentum tensor} of the point M,
while quantity $i^{0}_{[\mu\nu](\lambda)}$, specified in (3.11),
will be called {\it a geometrical momentum tensor} of the point M
relative to the "center of inertia" C.

\parindent=24pt\ Quantity
$$
  h_{\Hat{\alpha}\Hat{\beta}} = g_{\mu\nu} b^{\mu}_{\; \cdot
  \Hat{\alpha}} b^{\nu}_{\; \cdot \Hat{\beta}}\eqno{(3.35)}
$$
may be called {\it a deformation tensor}, forasmuch as it is
stipulated by moving of substance inside of the domain under
consideration. Definition (3.27) is somewhat differs from the
definition accepted in the mechanic of continuous media, where
deformation tensor is defined as semi-difference between external
background metrics and the metrics of deformed
body~\cite{Halb1}-\cite{Scho} either in the Lagrange-Green form or
in the Euler-Cauchy-Almansi form. However, the definition similar
to (3.27) is given, for example, in~\cite{Grot}, while
semi-difference above is called a strain tensor. As it will be
shown below, $h_{\Hat{\alpha}\Hat{\beta}}$ actually represents the
internal metric of the domain.

\parindent=24pt\ At last, if we introduce also following notations
$$
  \ell_{[\mu\nu]\Hat{\alpha}} = g_{\rho\sigma} b^{\rho}_{\; \cdot
  \Hat{\alpha}} \left [\eta^{\sigma(\tau)}
  i^{0}_{[\mu\nu](\tau)} - \Tilde{d}^{\sigma}_{\; \cdot [\mu\nu]}
  \right ] = e^{(\tau)}_{\; \cdot \rho} b^{\rho}_{\; \cdot
  \Hat{\alpha}} i_{[\mu\nu](\tau)} \; ,\eqno{(3.36)}
$$
$$
  c_{(\lambda)\Hat{\alpha}} = g_{\rho\sigma} e^{\rho}_{\; \cdot
  (\lambda)} b^{\sigma}_{\; \cdot \Hat{\alpha}} \; ,\eqno{(3.37)}
$$
the interval (2.11) will have the form
$$
  dS^{2} = \Sigma c^{2} d\tau^{2} = \eta_{(\mu)(\nu)} dx^{(\mu)}
  dx^{(\nu)} - \frac{1}{4}
  \Tilde{\Tilde{j}}_{[\rho\sigma][\tau\omega]}
  d\varphi^{[\rho\sigma]} d\varphi^{[\tau\omega]} +
  h_{\Hat{\alpha}\Hat{\beta}} dr^{\Hat{\alpha}} dr^{\Hat{\beta}} +
$$
$$
  + \Tilde{i}_{[\rho\sigma](\mu)} dx^{(\mu)}
  d\varphi^{[\rho\sigma]} + 2c_{(\mu)\Hat{\alpha}} dx^{(\mu)}
  dr^{\Hat{\alpha}} + \Tilde{\ell}_{[\rho\sigma]\Hat{\alpha}}
  dr^{\Hat{\alpha}} d\varphi^{[\rho\sigma]} \; ,\eqno{(3.38)}
$$
where
$$
 \Tilde{i}_{[\mu\nu](\lambda)} = \eta_{(\lambda)(\kappa)}
 e^{(\kappa)}_{\; \cdot \rho} \left [ \eta^{\rho(\sigma)}
 \Tilde{i}^{0}_{[\mu\nu](\sigma)} + d^{\rho}_{\; \cdot [\mu\nu]}
 \right ] \; ,\eqno{(3.39)}
$$
$$
 \Tilde{\ell}_{[\mu\nu]\Hat{\alpha}} = g_{\rho\sigma} b^{\rho}_{\;
 \cdot \Hat{\alpha}} \left [\eta^{\sigma(\tau)}
  \Tilde{i}^{0}_{[\mu\nu](\tau)} + d^{\sigma}_{\; \cdot [\mu\nu]}
  \right ] = e^{(\tau)}_{\; \cdot \rho} b^{\rho}_{\; \cdot
  \Hat{\alpha}} \Tilde{i}_{[\mu\nu](\tau)} \; ,\eqno{(3.40)}
$$
$\tau$ is proper time of the point M.

\parindent=24pt\ Proper time of the "center of inertia"
$\tau_{\mathrm{C}}$ may be defined from the relation
$$
 \sigma c^{2} d\tau^{2}_{\mathrm{C}} = \eta_{(\mu)(\nu)}
 dx^{(\mu)} dx^{(\nu)} \; ,\eqno{(3.41)}
$$
where $\sigma = \pm 1$ (obviously, a concept of proper time of the
"center of inertia" is undefinable for $\sigma = 0$). Then it
follows from (3.30) and (3.33) that
$$
 \frac{d\tau_{\mathrm{C}}}{d\tau} = \Gamma = \left [\Sigma \left (
 \sigma - \frac{1}{4} \Tilde{\Tilde{j}}_{[\rho\sigma][\tau\omega]}
 \Omega^{[\rho\sigma]}_{\mathrm{C}}
 \Omega^{[\tau\omega]}_{\mathrm{C}} + h_{\Hat{\alpha}\Hat{\beta}}
 V^{\Hat{\alpha}}_{\mathrm{C}} V^{\Hat{\beta}}_{\mathrm{C}} +
 \right. \right.
$$
$$
 \left. \left. + \Tilde{i}_{[\rho\sigma](\mu)}
 U^{(\mu)}_{\mathrm{C}} \Omega^{[\rho\sigma]}_{\mathrm{C}}
  + 2c_{(\mu)\Hat{\alpha}} U^{(\mu)}_{\mathrm{C}}
  V^{\Hat{\alpha}}_{\mathrm{C}} +
  \Tilde{\ell}_{[\rho\sigma]\Hat{\alpha}}
  V^{\Hat{\alpha}}_{\mathrm{C}} \Omega^{[\rho\sigma]}_{\mathrm{C}}
  \right ) \right ]^{-1/2} \; ,\eqno{(3.42)}
$$
where $U^{(\mu)}_{\mathrm{C}} = dx^{(\mu)}/cd\tau_{\mathrm{C}}$ is
four-velocity of translational movement of the medium in the
back-ground space, $\Omega^{[\rho\sigma]}_{\mathrm{C}} =
d\varphi^{[\rho\sigma]}/cd\tau_{\mathrm{C}}$ is angular velocity
of rotational movement of the medium relative to the "center of
inertia", $V^{\Hat{\alpha}}_{\mathrm{C}} =
dr^{\Hat{\alpha}}/cd\tau_{\mathrm{C}}$ is four-velocity of
translations of substance of the medium in the vicinity of the
point M relative to the "center of inertia".

\parindent=24pt\ The proper time of the point M will be coincided
with the proper time of the "center of inertia" when $\Gamma = 1$.
It is possible only in two cases: i) if internal substance of the
domain is absolutely rigid and performs only translational
movement, i.e. when $\Omega^{[\rho\sigma]} =
d\varphi^{[\rho\sigma]}/cd\tau = 0$, $V^{\Hat{\alpha}} \equiv
V^{\Hat{\alpha}}_{\mathrm{M}} = dr^{\Hat{\alpha}}/cd\tau = 0$; ii)
there takes place a relation
$$
 - \frac{1}{4} \Tilde{\Tilde{j}}_{[\rho\sigma][\tau\omega]}
 \Omega^{[\rho\sigma]}_{\mathrm{C}}
 \Omega^{[\tau\omega]}_{\mathrm{C}} + h_{\Hat{\alpha}\Hat{\beta}}
 V^{\Hat{\alpha}}_{\mathrm{C}} V^{\Hat{\beta}}_{\mathrm{C}} +
 + \Tilde{i}_{[\rho\sigma](\mu)} U^{(\mu)}_{\mathrm{C}}
 \Omega^{[\rho\sigma]}_{\mathrm{C}} +
$$
$$
 + 2c_{(\mu)\Hat{\alpha}} U^{(\mu)}_{\mathrm{C}}
  V^{\Hat{\alpha}}_{\mathrm{C}} +
  \Tilde{\ell}_{[\rho\sigma]\Hat{\alpha}}
  V^{\Hat{\alpha}}_{\mathrm{C}} \Omega^{[\rho\sigma]}_{\mathrm{C}}
  = 0 \; , \; \; \; \Sigma = \sigma \; .\eqno{(3.43)}
$$
If this relation is fulfilled for all points of the domain
occupied by the extended object one may say about proper time of
the object.

\parindent=24pt\ Expression (3.30) is turn out to be an interval
in 14-dimensional space, covered by coordinates $X^{A}$ ($A =
(\mu), [\rho\sigma], \Hat{\alpha}$) with
$$
  X^{(\mu)} = x^{(\mu)} \; , \; X^{[\rho\sigma]} =
  \varphi^{[\rho\sigma]} \; , \; X^{\Hat{\alpha}} =
  r^{\Hat{\alpha}} \; .\eqno{(3.44)}
$$
Then (3.30) may be written as
$$
  dS^{2} = G_{AB} dX^{A} dX^{B} \; .\eqno{(3.45)}
$$
The interval (3.36) obtained in such a way, generally speaking,
describes not necessarily extended particle, but any deformable
medium performing translational and rotational movement. Moreover
it is supposed, that the equations of motion of any point M of
this medium look like geodesic equation in 14-dimensional space
$$
 \dot{U}^{A} + \Gamma^{A}_{\; \cdot BC} U^{B} U^{C} = 0 \;
 ,\eqno{(3.46)}
$$
where
$$
 \Gamma^{A}_{\; \cdot BC} = \frac{1}{2} G^{AD} \left (
 \partial_{B} G_{DC} + \partial_{C} G_{BD} - \partial_{D} G_{BC}
 \right ) \; ,\eqno{(3.47)}
$$
$$
 U^{(\mu)} = \frac{dx^{(\mu)}}{cd\tau} = \Gamma
 U^{(\mu)}_{\mathrm{C}} \; ,\eqno{(3.48)}
$$
$$
 U^{[\rho\sigma]} = \Omega^{[\rho\sigma]} =
 \frac{d\varphi^{[\rho\sigma]}}{cd\tau} = \Gamma
 \Omega^{[\rho\sigma]}_{\mathrm{C}} \; ,\eqno{(3.49)}
$$
$$
 U^{\Hat{\alpha}} = V^{\Hat{\alpha}} =
 \frac{dr^{\Hat{\alpha}}}{cd\tau} = \Gamma
 V^{\Hat{\alpha}}_{\mathrm{C}} \; ,\eqno{(3.50)}
$$

\parindent=24pt\ To describe a rotating and deformable particle,
it is necessary to impose the certain conditions on metrics
$G_{AB}$, which are considered in the following paragraph.

\section{ Splitting of the equations of motion}

\parindent=24pt\ \; \; \; Let ${\bf R}_{14}$ denotes 14-dimensional
space, and ${\bf R}_{4}$, ${\bf H}_{6}$ and $\Hat{{\bf R}}_{4}$
denote the background space, covered by coordinates $x^{(\mu)}$, a
space of rotation and hyper-rotation angles
$\varphi^{[\rho\sigma]}$, and internal space, covered by
coordinates $r^{\Hat{\alpha}}$, respectively. Similarly, ${\bf
R}_{10}$, $\Hat{{\bf R}}_{10}$ and $\Hat{{\bf R}}_{8}$ be
denotations of spaces of coordinates $\{ x^{(\mu)},
\varphi^{[\rho\sigma]} \}$, $\{ r^{\Hat{\alpha}},
\varphi^{[\rho\sigma]} \}$ and $\{ x^{(\mu)}, r^{\Hat{\alpha}}
\}$, respectively. Thus, ${\bf R}_{4} = {\bf R}_{10} \bigcap
\Hat{{\bf R}}_{8}$, ${\bf H}_{6} = {\bf R}_{10} \bigcap \Hat{{\bf
R}}_{10}$, $\Hat{{\bf R}}_{4} = \Hat{{\bf R}}_{8} \bigcap
\Hat{{\bf R}}_{10}$. Obviously, it is possible to present ${\bf
R}_{14}$ in the form of the direct sum of spaces: ${\bf R}_{14} =
{\bf R}_{4} \oplus {\bf H}_{6} \oplus \Hat{{\bf R}}_{4}$ with
${\bf R}_{10} = {\bf R}_{4} \oplus {\bf H}_{6}$, $\Hat{{\bf
R}}_{10} = \Hat{{\bf R}}_{4} \oplus {\bf H}_{6}$, $\Hat{{\bf
R}}_{8} = {\bf R}_{4} \oplus \Hat{{\bf R}}_{4}$.

\parindent=24pt\ By definition, rotations and deformations
inside of a particle are independent movements, which are not
depending on movement in background space ${\bf R}_{4}$ and from
each other. Therefore, all three spaces, ${\bf R}_{4}$, ${\bf
H}_{6}$ and $\Hat{{\bf R}}_{4}$, should be in certain respects
independent and orthogonal to each other. This situation may be
described from a viewpoint of the embedded space theory. Actually,
the space ${\bf R}_{14}$ is trivial embedding space, so that ${\bf
R}_{4} \subset {\bf R}_{14}$, ${\bf H}_{6} \subset {\bf R}_{14}$,
$\Hat{{\bf R}}_{4} \subset {\bf R}_{14}$. Because for a free
particle we do not observe any dependence of external movement (in
${\bf R}_{4}$) on internal variables, the space ${\bf R}_{4}$
appears to be stationary hypersurface~\cite{Szek} in ${\bf
R}_{14}$ defined by conditions $\varphi^{[\rho\sigma]} = 0$,
$\Omega^{[\rho\sigma]} = 0$, $r^{\Hat{\alpha}} = 0$,
$V^{\Hat{\alpha}} = 0$. This leads to orthogonality of geodesics
of the space ${\bf R}_{4}$ to geodesics of space $\Hat{{\bf
R}}_{10}$, written as a condition
$$
 \left [ \frac{1}{2} \Tilde{i}_{[\rho\sigma](\mu)}
 \Omega^{[\rho\sigma]} + c_{(\mu)\Hat{\alpha}} V^{\Hat{\alpha}}
 \right ] U^{(\mu)} = 0 \; .\eqno{(4.1)}
$$
Then $(\mu)$-components of geodesic equations (3.35) in ${\bf
R}_{14}$ give geodesic equations in ${\bf R}_{4}$:
$$
 \dot{U}^{(\mu)} + \Gamma^{(\mu)}_{\; \cdot (\lambda)(\kappa)}
 U^{(\lambda)} U^{(\kappa)} = 0 \; ,\eqno{(4.2)}
$$
$$
 \nabla_{(\mu)} \Tilde{i}_{[\rho\sigma](\nu)} + \nabla_{(\nu)}
 \Tilde{i}_{[\rho\sigma](\mu)} = 2
 \frac{\partial\eta_{(\mu)(\nu)}}{\partial \varphi^{[\rho\sigma]}}
 = 0 \; ,\eqno{(4.3)}
$$
$$
 \nabla_{(\mu)} c_{(\nu)\Hat{\alpha}} + \nabla_{(\nu)}
 c_{(\mu)\Hat{\alpha}} =
 \frac{\partial\eta_{(\mu)(\nu)}}{\partial r^{\Hat{\alpha}}} = 0
 \; ,\eqno{(4.4)}
$$
where $\nabla_{(\mu)}$ denotes covariant derivative in ${\bf
R}_{4}$,
$$
 \Gamma^{(\mu)}_{\; \cdot (\lambda)(\kappa)} = \frac{1}{2}
 \eta^{(\mu)(\nu)}[ \partial_{(\lambda)} \eta_{(\nu)(\kappa)} +
 \partial_{(\kappa)} \eta_{(\lambda)(\nu)} - \partial_{(\nu)}
 \eta_{(\lambda)(\kappa)} ]\eqno{(4.5)}
$$
are coefficients of connection in ${\bf R}_{4}$, which ought to be
called an "external connection", where $\eta^{(\mu)(\nu)}$ satisfy
relations $\eta^{(\mu)(\lambda)} \eta_{(\lambda)(\nu)} =
\delta^{(\mu)}_{\; \cdot (\nu)}$.

\parindent=24pt\ The space $\Hat{{\bf R}}_{10} = \Hat{{\bf R}}_{4}
\oplus {\bf H}_{6}$ in turn is an embedding space for $\Hat{{\bf
R}}_{4}$ and ${\bf H}_{6}$. $\Hat{{\bf R}}_{4}$ is stationary
hypersurface in $\Hat{{\bf R}}_{10}$ defined by conditions
$\varphi^{[\rho\sigma]} = 0$, $\Omega^{[\rho\sigma]} = 0$. It
leads both to orthogonality of geodesics in $\Hat{{\bf R}}_{4}$
and ${\bf H}_{6}$,
$$
 \frac{1}{2} \Tilde{\ell}_{[\rho\sigma]\Hat{\alpha}}
 \Omega^{[\rho\sigma]} V^{\Hat{\alpha}} = 0 \; ,\eqno{(4.6)}
$$
and to equations of motion
$$
 \dot{V}^{\Hat{\alpha}} + \mathrm{H}^{\Hat{\alpha}}_{\; \cdot
 \Hat{\beta} \Hat{\gamma}} V^{\Hat{\beta}} U^{\Hat{\gamma}} = 0 \;
 ,\eqno{(4.7)}
$$
$$
 \nabla_{\Hat{\alpha}} c_{(\mu)\Hat{\beta}} + \nabla_{\Hat{\beta}}
 c_{(\mu)\Hat{\alpha}} = \partial_{(\mu)} h_{\Hat{\alpha}
 \Hat{\beta}} \; ,\eqno{(4.8)}
$$
$$
 \nabla_{\Hat{\alpha}} \Tilde{\ell}_{[\rho\sigma] \Hat{\beta}} +
 \nabla_{\Hat{\beta}} \Tilde{\ell}_{[\rho\sigma] \Hat{\alpha}} = 2
 \frac{\partial h_{\Hat{\alpha} \Hat{\beta}}}{\partial
 \varphi^{[\rho\sigma]}} = 0 \; ,\eqno{(4.9)}
$$
where $\nabla_{\Hat{\alpha}}$ denotes covariant derivative in
$\Hat{{\bf R}}_{4}$,
$$
 \mathrm{H}^{\Hat{\alpha}}_{\cdot \; \Hat{\beta} \Hat{\gamma}} =
 \frac{1}{2} h^{\Hat{\alpha} \Hat{\delta}} [
 \partial_{\Hat{\beta}} \eta_{\Hat{\delta}\Hat{\gamma}} +
 \partial_{\Hat{\gamma}} \eta_{\Hat{\beta}\Hat{\delta}} -
 \partial_{\Hat{\delta}} \eta_{\Hat{\beta}\Hat{\gamma}}
 ]\eqno{(4.10)}
$$
are coefficients of connection in $\Hat{{\bf R}}_{4}$, which ought
to be called an "internal connection" of extended particle;
$h^{\Hat{\alpha} \Hat{\beta}}$ satisfy relations $h^{\Hat{\alpha}
\Hat{\gamma}} h_{\Hat{\gamma} \Hat{\beta}} =
\delta^{\Hat{\alpha}}_{\cdot \; \Hat{\beta}}$ .

\parindent=24pt\ The space ${\bf R}_{14}$ may be represented also
as ${\bf R}_{14} = \Hat{{\bf R}}_{8} \oplus {\bf H}_{6}$ and
considered as embedding space for $\Hat{{\bf R}}_{8} \subset {\bf
R}_{14}$. Then $\Hat{{\bf R}}_{8}$ is stationary hypersurface in
${\bf R}_{14}$ defined by conditions $\varphi^{[\rho\sigma]} = 0$,
$\Omega^{[\rho\sigma]} = 0$. It leads both to orthogonality of
geodesics in $\Hat{{\bf R}}_{8}$ and ${\bf H}_{6}$ and to
equations of motion, which in view of (4.2)-(4.4), (4.7)-(4.9) are
led to the equations
$$
 \partial_{(\mu)} c_{(\nu)\Hat{\alpha}} - \partial_{(\nu)}
 c_{(\mu)\Hat{\alpha}} = \frac{\partial\eta_{(\mu)(\nu)}}{\partial
 r^{\Hat{\alpha}}} = 0 \; ,\eqno{(4.11)}
$$
$$
 \partial_{(\mu)} \Tilde{\ell}_{[\rho\sigma] \Hat{\alpha}} +
 \partial_{\Hat{\alpha}} \Tilde{i}_{[\rho\sigma](\mu)} = 2
 \frac{\partial c_{(\mu)\Hat{\alpha}}}{\partial
 \varphi^{[\rho\sigma]}} = 0 \; ,\eqno{(4.12)}
$$
$$
 \partial_{\Hat{\beta}} c_{(\mu)\Hat{\alpha}} -
 \partial_{\Hat{\alpha}} c_{(\mu)\Hat{\beta}} = \partial_{(\mu)}
 h_{\Hat{\alpha}\Hat{\beta}} = 0 \; .\eqno{(4.13)}
$$
Orthogonality condition (4.6) is complemented with two more
conditions
$$
 \frac{1}{2} \Tilde{i}_{[\rho\sigma](\mu)} \Omega^{[\rho\sigma]}
 U^{(\mu)} = 0 \; ,\eqno{(4.14)}
$$
$$
 c_{(\mu)\Hat{\alpha}} U^{(\mu)} V^{\Hat{\alpha}}= 0 \;
 .\eqno{(4.15)}
$$

\parindent=24pt\ In deriving equations (4.2)-(4.4), (4.7)-(4.9),
(4.11)-(4.13) we have been using an independence (4.3), (4.4) of
the background metric on coordinates $\varphi^{[\rho\sigma]}$ and
$r^{\Hat{\alpha}}$, as well as independence of the internal metric
of the space $\Hat{{\bf R}}_{4}$ on coordinates $x^{(\mu)}$ and
$\varphi^{[\rho\sigma]}$, following from equations (4.9) and
(4.13). Instead of (4.4), (4.8), (4.11) and (4.13) one may write
$$
 \nabla_{(\mu)} c_{(\nu)\Hat{\alpha}} = 0 \; , \;
 \nabla_{\Hat{\alpha}} c_{(\mu)\Hat{\beta}} = 0 \; .\eqno{(4.16)}
$$

\parindent=24pt\ In view of orthogonality conditions (4.6),
(4.14),(4.15) the relation (3.43) takes a simple form
$$
 -\frac{1}{4} \Tilde{\Tilde{j}}_{[\rho\sigma][\tau\omega]}
 \Omega^{[\rho\sigma]}_{\mathrm{C}}
 \Omega^{[\tau\omega]}_{\mathrm{C}} + h_{\Hat{\alpha}\Hat{\beta}}
 V^{\Hat{\alpha}}_{\mathrm{C}} V^{\Hat{\beta}}_{\mathrm{C}} = 0 \;
 , \; \; \; \Sigma = \sigma \; .\eqno{(4.17)}
$$

\parindent=24pt\ Let us clarify this relation in simplest case when i)
background space is the Minkowski space ${\bf R}_{4} \equiv {\bf
E}^{\mathrm{R}}_{1,3}$ with the metric $\eta =
\{\eta_{(\mu)(\nu)}\} = \mathrm{diag}{(+1,-1,-1,-1)}$, and ii)
l.h.s. of the equation (2.7) does not depend on angular
variations, i.e. $d^{\mu}_{\cdot \; [\lambda\kappa]} = 0$. In this
case we have $i_{[\mu\nu](\lambda)} = i^{0}_{[\mu\nu](\lambda)}$,
$j_{[\mu\nu][\lambda\kappa]} = j^{0}_{[\mu\nu][\lambda\kappa]}$.
In the reference frame of the "center of inertia" the motion of
the point M is describing by the deformation tensor
$h_{\Hat{\alpha}\Hat{\beta}}$, being an internal metric, defined
by a character of internal movements. On the other hand, in this
case $h_{\Hat{\alpha}\Hat{\beta}}$ should coincide with an
external metric $g_{\mu\nu}$, so that $b^{\mu}_{\cdot \;
\Hat{\alpha}} = \delta^{\mu}_{\cdot \; \Hat{\alpha}}$, and
relation (3.35) takes a form $h_{\Hat{\alpha}\Hat{\beta}} =
g_{\mu\nu} \delta^{\mu}_{\cdot \; \Hat{\alpha}}
\delta^{\nu}_{\cdot \; \Hat{\beta}}$, or $g_{\mu\nu} =
h_{\Hat{\alpha}\Hat{\beta}} \delta^{\Hat{\alpha}}_{\cdot \; \mu}
\delta^{\Hat{\beta}}_{\cdot \; \nu}$.

\parindent=24pt\ Let us introduce following notations
$$
  \Omega_{(k)} = \frac{1}{2} \varepsilon_{(k)(l)(m)}
  \Omega^{[lm]}_{\mathrm{C}} \; , \; \; \; B^{(k)} =
  \Omega^{[0k]}_{\mathrm{C}} \; .\eqno{(4.18)}
$$
Obviously, $\Omega_{(k)}$ is a component of pseudovector ${\bf
\Omega} = \{\Omega_{(k)}\}$ of angular velocity of rotation of the
object in the plane $(l,m)$, perpendicular to the axis $(k)$;
${\bf B} = \{B^{(k)}\}$ is some vector, whose physical meaning
will have to clarify yet. Then the equation (4.17) may be write
down in the form
$$
  -{[{\bf \Omega} \times {\bf r}]}^{2} - 2r^{\Hat{0}} ({\bf B}
  \cdot [{\bf \Omega} \times {\bf r}]) + (r^{\Hat{0}})^{2} {\bf
  B}^{2} - ({\bf r} \cdot {\bf B})^{2} +
  h_{\Hat{\alpha}\Hat{\beta}} V^{\Hat{\alpha}}_{\mathrm{C}}
  V^{\Hat{\beta}}_{\mathrm{C}} = 0 \; ,\eqno{(4.19)}
$$
where scalar and vector products are determined relative to
background metric:
$$
  {[{\bf \Omega} \times {\bf r}]}^{(k)} = \varepsilon^{(k)(l)(m)}
  \eta_{(m)\Hat{\alpha}} \Omega_{(l)} r^{\Hat{\alpha}} \; , \; \;
  \; {[{\bf \Omega} \times {\bf r}]}^{2} = {\bf r}^{2} {\bf
  \Omega}^{2} - ({\bf r} \cdot {\bf \Omega})^{2} \; ,\eqno{(4.20)}
$$
$$
  ({\bf B} \cdot [{\bf \Omega} \times {\bf r}]) = -\eta_{(m)(k)}
  B^{(m)} [{\bf \Omega} \times {\bf r}]^{(k)} \; ,\eqno{(4.21)}
$$
$$
  {\bf r}^{2} = -\eta_{(m)(n)} \delta^{(m)}_{ \; \cdot \;
  \Hat{\alpha}} \delta^{(n)}_{ \; \cdot \; \Hat{\beta}}
  r^{\Hat{\alpha}} r^{\Hat{\beta}} \; , \; \; {\bf B}^{2} =
  -\eta_{(m)(n)} B^{(m)} B^{(n)} \; , \; \; {\bf \Omega}^{2} =
  -\eta^{(m)(n)} \Omega_{(m)} \Omega_{(n)} \; ,\eqno{(4.22)}
$$
$$
  ({\bf r} \cdot {\bf \Omega}) = \delta^{(m)}_{ \; \cdot \;
  \Hat{\alpha}} r^{\Hat{\alpha}} \Omega_{(m)} \; , \; \; ({\bf r}
  \cdot {\bf B}) = -\eta_{(m)\Hat{\alpha}} B^{(m)}
  r^{\Hat{\alpha}} \; .\eqno{(4.23)}
$$

\parindent=24pt\ In the reference frame of the "center of inertia"
the orthogonality condition (4.6) looks as follows
$$
  \frac{1}{2} \Tilde{\ell}_{[\mu\nu]\Hat{\alpha}}
  \Omega^{[\mu\nu]} V^{\Hat{\alpha}}_{\mathrm{C}} =
  [e^{(0)}_{ \; \cdot \; \alpha} ({\bf r} \cdot {\bf B}) -
  r^{\Hat{0}} ({\bf e}_{\alpha} \cdot {\bf B}) - ({\bf e}_{\alpha}
  \cdot [{\bf \Omega} \times {\bf r}])]
  V^{\Hat{\alpha}}_{\mathrm{C}} = 0 \; ,\eqno{(4.24)}
$$
where vectors ${\bf e}_{\alpha} = \{e^{(m)}_{ \; \cdot \;
\alpha}\}$ are spatial components of moving 4-hedron and
$$
  ({\bf e}_{\alpha} \cdot {\bf B}) = -\eta_{(m)(n)}
  e^{(m)}_{ \; \cdot \; \Hat{\alpha}} B^{(n)} \; , \; \; ({\bf
  e}_{\alpha} \cdot [{\bf \Omega} \times {\bf r}]) =
  -\eta_{(m)(n)} e^{(m)}_{ \; \cdot \; \Hat{\alpha}} [{\bf \Omega}
  \times {\bf r}]^{(n)} \; .\eqno{(4.25)}
$$

\parindent=24pt\ Conditions (4.14) and (4.15) in arbitrary
reference frame look as
$$
  \frac{1}{2} \Tilde{i}_{[\mu\nu](\lambda)} \Omega^{[\mu\nu]}
  U^{(\lambda)} =
  V^{\Hat{0}}_{\mathrm{C}} ({\bf r} \cdot {\bf B}) - r^{\Hat{0}}
  ({\bf V}_{\mathrm{C}} \cdot {\bf B}) - ({\bf V}_{\mathrm{C}}
  \cdot [{\bf \Omega} \times {\bf r}]) = 0 \; ,\eqno{(4.26)}
$$
$$
  c_{(\mu)\Hat{\alpha}} U^{(\mu)} V^{\Hat{\alpha}}_{\mathrm{C}} =
  c_{(0)\Hat{\alpha}} U^{(0)} V^{\Hat{\alpha}}_{\mathrm{C}} +
  c_{(m)\Hat{\alpha}} U^{(m)} V^{\Hat{\alpha}}_{\mathrm{C}} = 0 \;
  .\eqno{(4.27)}
$$
In the reference frame of the "center of inertia", where $U^{(0)}
= 1$, $U^{(m)} = 0$, conditions (4.24) (4.26), (4.27) are reduced
to
$$
  ({\bf V}_{\mathrm{C}} \cdot \{- r^{\Hat{0}} {\bf B} + [{\bf
  \Omega} \times {\bf r}] \}) = 0 \; ,\eqno{(4.28)}
$$
$$
  ({\bf r} \cdot {\bf B}) = 0 \; ,\eqno{(4.29)}
$$
$$
  h_{\Hat{\alpha}\Hat{\beta}} e^{\Hat{\alpha}}_{ \; \cdot \; (0)}
  V^{\Hat{\beta}}_{\mathrm{C}} = 0 \; .\eqno{(4.30)}
$$
where
$$
  {\mathbf{V}}_{\mathrm{C}} = V^{\Hat{\alpha}}_{\mathrm{C}}
  \mathbf{e}_{\alpha} = \{ V^{(m)}_{\mathrm{C}} \} = \{
  V^{\Hat{\alpha}}_{\mathrm{C}} e^{(m)}_{ \; \cdot \; \alpha} \}
  \; .\eqno{(4.31)}
$$

\parindent=24pt\ It follows from (4.28), that
$$
  r^{\Hat{0}} {\bf B} = [{\bf \Omega} \times {\bf r}] + [{\bf
  V}_{\mathrm{C}} \times {\bf K}] \; .\eqno{(4.32)}
$$
As well it follows from (4.29) and (4.32), that $({\bf r} \cdot
[{\bf V}_{\mathrm{C}} \times {\bf K}]) = ({\bf K} \cdot [{\bf r}
\times {\bf V}_{\mathrm{C}}]) = 0$. Hence, pseudovector $\bf K$,
perpendicular to the vector ${\bf V}_{\mathrm{C}}$ and lying in
the plane formed by the vectors $\bf r$ and ${\bf
V}_{\mathrm{C}}$, equals to ${\bf K} = a [{\bf V}_{\mathrm{C}}
\times [{\bf r} \times {\bf V}_{\mathrm{C}}]]$, whence
$$
  r^{\Hat{0}} {\bf B} = [\{{\bf \Omega} + a{\bf V}^{2}_{\mathrm{C}}
  {\bf V}_{\mathrm{C}} \} \times {\bf r}] \; .\eqno{(4.33)}
$$
If pseudoscalar $a = 0$, then the vector ${\bf B}$ is parallel to
the vector $[{\bf \Omega} \times {\bf r}]$ of linear velocity of
the motion of the point M around the axis ${\bf \Omega}$ due to
only rotation of extended object, and condition (4.29) is
fulfilled automatically. In view of (4.33) the condition (4.19) in
the case above takes the form
$$
  a^{2} {\bf V}^{4}_{\mathrm{C}} [{\bf r} \times {\bf
  V}_{\mathrm{C}}]^{2} - 2 [{\bf \Omega} \times {\bf r}]^{2} +
  h_{\Hat{\alpha}\Hat{\beta}} V^{\Hat{\alpha}}_{\mathrm{C}}
  V^{\Hat{\beta}}_{\mathrm{C}} = 0 \; .\eqno{(4.34)}
$$

\parindent=24pt\ Orthogonality conditions (4.6), (4.14) and (4.15)
imply, that equations of motion (3.46) in the space ${\bf R}_{14}$
are splitted in equations (4.2), (4.3), (4.7), (4.9), (4.12) and
(4.16). These equations may be written down in the other form by
the introduction of the generalized momenta, conjugated with
coordinates $x^{(\mu)}$, $r^{\Hat{\alpha}}$,
$\varphi^{[\rho\sigma]}$.

\parindent=24pt\ As it is well known, equations (3.46), determining
a motion of point M, one can obtain from the action, representable
similar to the relativistic mass point action in the form
$$
 J = \int L c d \tau = -c \int [\Sigma dS^{2}]^{1/2} \;
 ,\eqno{(4.35)}
$$
where $dS^{2}$ is defined in (3.30), and the mass of the point is
supposed to be unit.

\parindent=24pt\ We define the specific generalized momenta
$$
 p_{(\mu)} = -\frac{\partial L}{c \partial U^{(\mu)}} = \Sigma c
 \left [\eta_{(\mu)(\nu)} U^{(\nu)} + \frac{1}{2}
 \Tilde{i}_{[\rho\sigma](\mu)} \Omega^{[\rho\sigma]} +
 c_{(\mu)\Hat{\alpha}} V^{\Hat{\alpha}} \right ] = \hspace{10mm}
$$
$$
 = p^{\mathrm{C}}_{(\mu)} + p^{\mathrm{rot}}_{(\mu)} +
 p^{\mathrm{M}}_{(\mu)}\; ;\hspace{27mm}\eqno{(4.36)}
$$
$$
 s_{[\lambda\kappa]} = -\frac{\partial L}{c \partial
 \Omega^{[\lambda\kappa]}} = \Sigma c \left [-\frac{1}{4}
 \Tilde{\Tilde{j}}_{[\lambda\kappa][\rho\sigma]}
 \Omega^{[\rho\sigma]} + \frac{1}{2}
 \Tilde{i}_{[\lambda\kappa](\nu)} U^{(\nu)} + \frac{1}{2}
 \Tilde{\ell}_{[\lambda\kappa]\Hat{\alpha}} V^{\Hat{\alpha}}
 \right ] = \hspace{1mm}
$$
$$
 = \frac{1}{2} \Tilde{i}_{[\lambda\kappa](\mu)} p_{(\nu)}
 = s^{\mathrm{rot}}_{[\lambda\kappa]} +
 s^{\mathrm{C}}_{[\lambda\kappa]} +
 s^{\mathrm{M}}_{[\lambda\kappa]} \; ,\eqno{(4.37)}
$$
$$
 \pi_{\Hat{\alpha}} = -\frac{\partial L}{c \partial
 V^{\Hat{\alpha}}} = \Sigma c \left [h_{\Hat{\alpha}\Hat{\beta}}
 V^{\Hat{\beta}} + c_{(\nu)\Hat{\alpha}} U^{(\nu)} + \frac{1}{2}
 \Tilde{\ell}_{[\rho\sigma]\Hat{\alpha}} \Omega^{[\rho\sigma]}
 \right ] = \hspace{13mm}
$$
$$
 = c_{(\mu)\Hat{\alpha}} \eta^{(\mu)(\nu)} p_{(\nu)} =
 \pi^{\mathrm{M}}_{\Hat{\alpha}} + \pi^{\mathrm{C}}_{\Hat{\alpha}}
 + \pi^{\mathrm{M}}_{\Hat{\alpha}} \; ,\hspace{2mm}\eqno{(4.38)}
$$
and generalized forces $f_{A} = - \partial L / \partial X^{A}$.
Then in view of (4.2)-(4.18) equations (3.35) are splitted in
system of the following equations
$$
 \frac{D_{\eta} p^{\mathrm{C}}_{(\mu)}}{cd\tau} =
 \frac{dp^{\mathrm{C}}_{(\mu)}}{cd\tau} - \Gamma^{(\rho)}_{ \;
 \cdot \; (\mu)(\nu)} U^{(\nu)} p^{\mathrm{C}}_{(\rho)} = 0 \;
 ,\hspace{25mm}\eqno{(4.39)}
$$
$$
 \frac{D_{\eta} p^{\mathrm{rot}}_{(\mu)}}{cd\tau} = \frac{\Sigma
 c}{16} \left [ \frac{\partial
 {\Tilde{i}}_{[\lambda\kappa](\mu)}}{\partial
 \varphi^{[\rho\sigma]}} - \frac{1}{2} \partial_{(\mu)}
 {\Tilde{\Tilde{j}}}_{[\lambda\kappa][\rho\sigma]} \right ]
 \Omega^{[\lambda\kappa]} \Omega^{[\rho\sigma]} \;
 ,\hspace{6mm}\eqno{(4.40)}
$$
$$
 \frac{D_{\eta} p^{\mathrm{M}}_{(\mu)}}{cd\tau} = \frac{\Sigma
 c}{2} \left [ {\Tilde{i}}_{[\lambda\kappa](\mu)}
 \dot{\Omega}^{[\lambda\kappa]} - \frac{1}{8} \partial_{(\mu)}
 {\Tilde{\Tilde{j}}}_{[\lambda\kappa][\rho\sigma]}
 \Omega^{[\lambda\kappa]} \Omega^{[\rho\sigma]} \right ] =
$$
$$
 = \frac{\Sigma c}{2} {\Tilde{i}}_{[\lambda\kappa](\nu)} \left
 [\delta^{(\nu)}_{\; \cdot \; (\mu)}
 \dot{\Omega}^{[\lambda\kappa]} + \frac{1}{4} \eta^{(\nu)(\tau)}
 \nabla_{(\mu)} {\Tilde{i}}_{[\rho\sigma](\tau)}
 \Omega^{[\lambda\kappa]} \Omega^{[\rho\sigma]} \right ] \; ;
 \hspace{-29mm}\eqno{(4.41)}
$$
$$
 \frac{D_{h} \pi^{\mathrm{M}}_{\Hat{\alpha}}}{cd\tau} =
 \frac{d \pi^{\mathrm{M}}_{\Hat{\alpha}}}{cd\tau} -
 H^{\Hat{\gamma}}_{ \; \cdot \; \Hat{\alpha}\Hat{\beta}}
 V^{\Hat{\beta}} \pi^{\mathrm{M}}_{\Hat{\gamma}} = 0 \;
 ,\hspace{33mm}\eqno{(4.42)}
$$
$$
 \frac{D_{h} \pi^{\mathrm{rot}}_{\Hat{\alpha}}}{cd\tau} =
 \frac{\Sigma c}{8} \left [ \frac{\partial
 {\Tilde{\ell}}_{[\lambda\kappa]\Hat{\alpha}}}{\partial
 \varphi^{[\rho\sigma]}} - \frac{1}{4} \partial_{\Hat{\alpha}}
 {\Tilde{\Tilde{j}}}_{[\lambda\kappa][\rho\sigma]} \right ]
 \Omega^{[\lambda\kappa]} \Omega^{[\rho\sigma]} \;
 ,\hspace{11mm}\eqno{(4.43)}
$$
$$
 \frac{D_{h} \pi^{\mathrm{M}}_{\Hat{\alpha}}}{cd\tau} =
 \frac{\Sigma c}{2} \left [
 {\Tilde{\ell}}_{[\lambda\kappa]\Hat{\alpha}}
 \dot{\Omega}^{[\lambda\kappa]} - \frac{1}{8}
 \partial_{\Hat{\alpha}}
 {\Tilde{\Tilde{j}}}_{[\lambda\kappa][\rho\sigma]}
 \Omega^{[\lambda\kappa]} \Omega^{[\rho\sigma]} \right ] =
 \hspace{4mm}
$$
$$
 = \frac{\Sigma c}{2}
 {\Tilde{\ell}}_{[\lambda\kappa]\Hat{\beta}} \left
 [\delta^{\Hat{\beta}}_{\; \cdot \; \Hat{\alpha}}
 \dot{\Omega}^{[\lambda\kappa]} + \frac{1}{4}
 h^{\Hat{\beta}\Hat{\gamma}} \nabla_{\Hat{\alpha}}
 {\Tilde{\ell}}_{[\rho\sigma]\Hat{\gamma}}
 \Omega^{[\lambda\kappa]} \Omega^{[\rho\sigma]} \right ] \; ;
 \hspace{-17mm}\eqno{(4.44)}
$$
$$
 \frac{ds_{[\mu\nu]}}{cd\tau} = \frac{1}{2}
 \eta^{(\lambda)(\kappa)} p_{(\lambda)} \frac{\partial
 {\Tilde{i}}_{[\rho\sigma](\kappa)}}{\partial \varphi^{[\mu\nu]}}
 \Omega^{[\rho\sigma]} \; ,\hspace{30mm}\eqno{(4.45)}
$$
$$
 \frac{ds^{\mathrm{rot}}_{[\mu\nu]}}{cd\tau} = 2 \frac{d(
 s^{\mathrm{C}}_{[\mu\nu]} +
 s^{\mathrm{M}}_{[\mu\nu]})}{cd\tau} + \frac{\Sigma c}{2}
 \Tilde{\Tilde{j}}_{[\mu\nu][\rho\sigma]}
 \dot{\Omega}^{[\rho\sigma]} - \hspace{13mm}
$$
$$
 - \eta^{(\lambda)(\kappa)}
 (p^{\mathrm{C}}_{(\lambda)} + p^{\mathrm{M}}_{(\lambda)})
 \frac{\partial \Tilde{i}_{[\mu\nu](\kappa)}}{\partial
 \varphi^{[\rho\sigma]}} \Omega^{[\rho\sigma]} \;
 .\hspace{2mm}\eqno{(4.46)}
$$

\parindent=24pt\ Orthogonality conditions (4.6), (4.14), (4.15)
reduce the interval (3.38) to the form
$$
 dS^{2} = \Sigma c^{2} d\tau^{2} = \eta_{(\mu)(\nu)} dx^{(\mu)}
 dx^{(\nu)} - \frac{1}{4}
 \Tilde{\Tilde{j}}_{[\lambda\kappa][\rho\sigma]}
 d\varphi^{[\lambda\kappa]} d\varphi^{[\rho\sigma]} +
 h_{\Hat{\alpha}\Hat{\beta}} dr^{\Hat{\alpha}}
 dr^{\Hat{\beta}}\eqno{(4.47)}
$$
with additional condition that
$\Tilde{\Tilde{j}}_{[\lambda\kappa][\rho\sigma]}
 \Omega^{[\lambda\kappa]} \Omega^{[\rho\sigma]}$ does not depend
 on $x^{(\mu)}$, but, generally speaking, may depend on
 $\varphi^{[\lambda\kappa]}$. Starting from (4.3), (4.9),
(4.12) and (4.16) one may show, that
$$
 \partial_{\Hat{\alpha}} \Tilde{i}_{[\rho\sigma](\mu)} =
 \eta^{(\nu)(\tau)} c_{(\nu)\Hat{\alpha}} \nabla_{(\tau)}
 \Tilde{i}_{[\rho\sigma](\mu)} \; ,\eqno{(4.48)}
$$
$$
 \partial_{\Hat{\alpha}}
 \Tilde{\Tilde{j}}_{[\lambda\kappa][\rho\sigma]} =
 \eta^{(\nu)(\tau)} c_{(\nu)\Hat{\alpha}} \partial_{(\tau)}
 \Tilde{\Tilde{j}}_{[\rho\sigma][\lambda\kappa]} \;
 .\hspace{2mm}\eqno{(4.49)}
$$

\parindent=24pt\ Now, independent conditions (4.3), (4.12) and
(4.48) are remained instead of interdependent conditions (4.3),
(4.9), (4.12) and (4.16). All the other relations (4.4), (4.8),
(4.9), (4.11)-(4.13), (4.16) and (4.49) are their consequences.
Interdependency of these relations is result of {\it degeneracy}
of the metric $G_{AB}$. It is manifested also in connection
(4.37), (4.38) between $p_{(\mu)}$, $s_{[\lambda\kappa]}$ and
$\pi_{\Hat{\alpha}}$. Owing to this fact equations of motion
(4.39)-(4.46) reduce to only three ones: (4.39), (4.42) and
$$
 \frac{ds^{\mathrm{rot}}_{[\mu\nu]}}{cd\tau} = - \frac{\Sigma
 c}{8} \frac{\partial
 \Tilde{\Tilde{j}}_{[\lambda\kappa][\rho\sigma]}}{\partial
 \varphi^{[\mu\nu]}} \Omega^{[\lambda\kappa]}
 \Omega^{[\rho\sigma]} \; .\eqno{(4.50)}
$$

\parindent=24pt\ Quantity $p^{\mathrm{C}}_{(\mu)}$ is a specific
momentum of the "center of inertia",
$\pi^{\mathrm{M}}_{\Hat{\alpha}}$ is a momentum of point M
relative to the "center of inertia", caused by movement of
substance inside extended object, $s^{\mathrm{rot}}_{[\mu\nu]}$ is
angular momentum of point M relative to the "center of inertia",
caused by rotation of this object. As to dependence of the inertia
tensor on $\varphi^{[\mu\nu]}$ it is reasonable to require it to
be invariant under translations on angular variables
$\varphi^{[mn]}$, $m, n = 1,2,3$. Because of an obscurity of
variables $\varphi^{[0n]}$ it would be necessary, generally
speaking, to conserve some freedom in dependence of the inertia
tensor on these variables. However, linking
$s^{\mathrm{rot}}_{[\mu\nu]}$ with spin of the particle (which
should be determined individually) and taking into account the
observable fact, that spin of free particles conserves its value,
we shall require a translational invariance on $\varphi^{[0n]}$ as
well. It leads to the equation
$ds^{\mathrm{rot}}_{[\mu\nu]}/cd\tau = 0$, which means
$\Omega^{[\mu\nu]} = \mathrm{const}$.

\parindent=24pt\ Thus, we have gave one of possible variants of
the description of free relativistic extended particle, where it
is necessary to start from the equations of motion (4.39), (4.42)
and (4.50). If the background space is the Minkowski space, then
conservation of spin (or angular velocity of rotation) as well as
conservation of momentum of extended particle take place as
consequences of these equations. Even in this case the internal
space is curved.

\section{ Degeneracy of the metric}

\parindent=24pt\ \; \; \; Let us summarize the results obtained
above. Firstly, representation of the space-time interval (3.38)
of the world line of a point M inside of a particle in
14-dimensional space, covered by the center-of-inertia, angular
and deformational coordinates, in the form (4.47) is possible
provided imposing the orthogonality conditions (4.6), (4.14),
(4.15). This is equivalent to splitting of ${\bf R}_{14}$ in three
subspaces ${\bf R}_{4}$, ${\bf H}_{6}$ and $\Hat{{\bf R}}_{4}$. In
this case the general equation of motion (3.46) also splits in
equation (4.2) (or (4.39)) of center-of-inertia motion, equation
(4.7) (or (4.42)) of motion of the point M in $\Hat{{\bf R}}_{4}$
relative to the center of inertia, which does not depend on
external motions, and equation (4.50) of rotational motion of the
point M around the center of inertia. Secondly, the interval
(4.47) along world line of a point M in ${\bf R}_{14}$ turns out
to be equivalent to the interval (3.41) along center-of-inertia
world line in ${\bf R}_{4}$ provided fulfillment of condition
(4.17). This fact makes possible to introduce time coinciding with
proper time of the center of inertia in all points of the region
in ${\bf R}_{4}$, occupied by an extended object. Thirdly,
equivalence of intervals (4.47) and (3.41) means the metric
$G_{AB}$ of the space ${\bf R}_{14}$ be degenerate and have rank
which is equal to the rank of background metric.

\parindent=24pt\ Obtained results are compatible with the theory
of degenerate Riemannian spaces~\cite{Jan}, according to which the
space ${\bf R}_{14}$ is reducible and there take place relations:
$$
 \partial_{\Hat{\alpha}} \eta_{(\mu)(\nu)} = 0 \; , \; \;
 \partial_{[\lambda\kappa]} \eta_{(\mu)(\nu)} = 0 \; , \; \;
 \partial_{[\lambda\kappa]} h_{\Hat{\alpha}\Hat{\beta}} = 0 \;
 ;\eqno{(5.1)}
$$
$$
 \nabla_{(\mu)} c_{(\nu)\Hat{\alpha}} = 0 \; , \; \;
 \nabla_{\Hat{\alpha}} c_{(\mu)\Hat{\beta}} = 0 \; , \; \;
 \partial_{[\lambda\kappa]} c_{(\mu)\Hat{\alpha}} = 0 \;
 ;\eqno{(5.2)}
$$
$$
 \nabla_{(\mu)} \Tilde{i}_{(\nu)[\lambda\kappa]} = 0 \; , \; \;
 \partial_{\Hat{\alpha}} \Tilde{i}_{(\mu)[\lambda\kappa]} = 0 \; ,
 \; \; \partial_{[\lambda\kappa]} \Tilde{i}_{(\mu)[\rho\sigma]} =
 0 \; ;\eqno{(5.3)}
$$
$$
 \partial_{(\mu)} \Tilde{\ell}_{\Hat{\alpha}[\lambda\kappa]} =
 \partial_{(\mu)} \Tilde{\ell}_{[\lambda\kappa]\Hat{\alpha}} = 0
 \; , \; \; \nabla_{\Hat{\beta}}
 \Tilde{\ell}_{\Hat{\alpha}[\lambda\kappa]} = 0 \; , \; \;
 \partial_{[\lambda\kappa]}
 \Tilde{\ell}_{\Hat{\alpha}[\rho\sigma]} = 0 \; ;\eqno{(5.4)}
$$
$$
 \partial_{(\mu)} h_{\Hat{\alpha}\Hat{\beta}} = 0 \; , \; \;
 \partial_{(\mu)} \Tilde{\Tilde{j}}_{[\lambda\kappa][\rho\sigma]}
 = 0 \; , \; \; \partial_{\Hat{\alpha}}
 \Tilde{\Tilde{j}}_{[\lambda\kappa][\rho\sigma]} = 0 \;
 .\eqno{(5.5)}
$$
Relations (5.1)-(5.3) are independent and compatible with both
relations (4.3), (4.16) and equations of motion (4.39), (4.42) and
(4.50). However, it follows from the last equation in (5.5) that
even for absolutely rigid body, when inertia tensor has the form
(3.1) (or (3.15)), $j^{0}_{[\lambda\kappa][\rho\sigma]} = 0$ due
to identity
$$
 r^{\Hat{\alpha}} \partial_{\Hat{\alpha}} (r^{2}_{\eta}
 D^{(\mu)(\nu)}) = 2 r^{2}_{\eta} D^{(\mu)(\nu)} \; ,\eqno{(5.6)}
$$
whence
$$
 r^{\Hat{\alpha}} \partial_{\Hat{\alpha}}
 j^{0}_{[\lambda\kappa][\rho\sigma]} = 2
 j^{0}_{[\lambda\kappa][\rho\sigma]} \; , \; \;
 r^{\Hat{\alpha}} \partial_{\Hat{\alpha}}
 \Tilde{\Tilde{j}}^{0}_{[\lambda\kappa][\rho\sigma]} = 2
 \Tilde{\Tilde{j}}^{0}_{[\lambda\kappa][\rho\sigma]} \;
 .\eqno{(5.7)}
$$
In the general case condition $\partial_{\Hat{\alpha}}
 \Tilde{\Tilde{j}}_{[\lambda\kappa][\rho\sigma]} = 0$ (more
 exactly, two conditions, $\nabla_{\Hat{\alpha}}
 c_{(\mu)\Hat{\beta}} = 0$ and $\partial_{\Hat{\alpha}}
 \Tilde{i}_{(\mu)[\lambda\kappa]} = 0$) expresses restriction to
 movement of substance inside of an extended object, and it is not
 necessarily this condition leads to
 $j_{[\lambda\kappa][\rho\sigma]} = 0$.

\parindent=24pt\ It follows from the last condition in (5.3) that
$\partial_{[\mu\nu]}
\Tilde{\Tilde{j}}_{[\lambda\kappa][\rho\sigma]} = 0$. Then
equation of motion (4.50) means, that all internal points are
rotating with the same constant angular velocity. In the general
case it is not quite necessarily. Therefore condition
$\partial_{[\mu\nu]} \Tilde{i}_{(\mu)[\rho\sigma]} = 0$, as well
as some other conditions (5.1)-(5.3), is not necessary too.

\parindent=24pt\ It should be noted that the main result of
splitting of equation of motion have to be in geodesic motion of
the center of inertia in the background space ${\bf R}_{4}$ and
geodesic motion of the point M in the internal space ${\Hat{\bf
R}}_{4}$ provided the rotation is absent. Moreover, total proper
angular momentum, which may be defined as
$$
 S_{[\mu\nu]} = {\mathop{\int}\limits_{V}} \rho
 s^{\mathrm{rot}}_{[\mu\nu]} dV \; ,\eqno{(5.8)}
$$
where $\rho$ is mass density, $V$ is volume of the region,
occupied by extended object, should be conserved. Obviously, it
follows from $j_{[\lambda\kappa][\rho\sigma]} = 0$ that
$S_{[\mu\nu]} = 0$. Such an object may be associated with scalar
particle.

\parindent=24pt\ We shall consider more adequate model, in which
${\bf R}_{14}$ is trivial embedding space for ${\bf R}_{4}$ and
${\Hat{\bf R}}_{10}$, but ${\Hat{\bf R}}_{10}$ is not an embedding
space for ${\Hat{\bf R}}_{4}$ and ${\bf H}_{6}$. In this case
coefficients of connection $\Gamma^{\Hat{\alpha}}_{\; \cdot \;
\Hat{\beta}[\lambda\kappa]}$, $\Gamma^{[\rho\sigma]}_{\; \cdot \;
\Hat{\beta}[\lambda\kappa]}$, $\Gamma^{\Hat{\alpha}}_{\; \cdot \;
[\rho\sigma][\lambda\kappa]}$, $\Gamma^{[\lambda\kappa]}_{\; \cdot
\; \Hat{\alpha}\Hat{\beta}}$ are, in general, not equal to zero
and instead of conditions (5.1)-(5.5) there will be fulfilled
conditions~\cite{Jan}
$$
 \partial_{\Hat{\alpha}} \eta_{(\mu)(\nu)} = 0 \; , \; \;
 \partial_{[\lambda\kappa]} \eta_{(\mu)(\nu)} = 0 \; ;\eqno{(5.9)}
$$
$$
 \nabla_{(\mu)} c_{(\nu)\Hat{\alpha}} = \partial_{(\mu)}
 c_{(\nu)\Hat{\alpha}} - \Gamma^{(\lambda)}_{ \; \cdot \;
 (\mu)(\nu)} c_{(\lambda)\Hat{\alpha}} = 0 \; ,\eqno{(5.10)}
$$
$$
 \nabla_{\Hat{\alpha}} c_{(\mu)\Hat{\beta}} =
 \partial_{\Hat{\alpha}} c_{(\mu)\Hat{\beta}} - H^{\Hat{\gamma}}_{
 \; \cdot \; \Hat{\alpha}\Hat{\beta}} c_{(\mu)\Hat{\gamma}} -
 \frac{1}{2} \Gamma^{[\lambda\kappa]}_{ \; \cdot \;
 \Hat{\alpha}\Hat{\beta}} \Tilde{i}_{(\mu)[\lambda\kappa]} = 0 \;
 ;\eqno{(5.11)}
$$
$$
 \nabla_{[\lambda\kappa]} c_{(\mu)\Hat{\alpha}} =
 \partial_{[\lambda\kappa]} c_{(\mu)\Hat{\alpha}} -
 \Gamma^{\Hat{\gamma}}_{ \; \cdot \; [\lambda\kappa]\Hat{\alpha}}
 c_{(\mu)\Hat{\gamma}} - \frac{1}{2} \Gamma^{[\rho\sigma]}_{ \;
 \cdot \; [\lambda\kappa]\Hat{\alpha}}
 \Tilde{i}_{(\mu)[\rho\sigma]} = 0 \; ;\eqno{(5.12)}
$$
$$
 \nabla_{(\mu)} \Tilde{i}_{(\nu)[\lambda\kappa]} =
 \partial_{(\mu)} \Tilde{i}_{(\nu)[\lambda\kappa]} -
 \Gamma^{\rho}_{ \; \cdot \; \mu\nu}
 \Tilde{i}_{(\rho)[\lambda\kappa]} = 0 \; ,\eqno{(5.13)}
$$
$$
 \nabla_{\Hat{\alpha}} \Tilde{i}_{(\mu)[\lambda\kappa]} =
 \partial_{\Hat{\alpha}} \Tilde{i}_{(\mu)[\lambda\kappa]} -
 2 \Gamma^{\Hat{\gamma}}_{ \; \cdot \;
 \Hat{\alpha}[\lambda\kappa]} c_{(\mu)\Hat{\gamma}} -
 \Gamma^{[\rho\sigma]}_{ \; \; \cdot \; \;
 \Hat{\alpha}[\lambda\kappa]} \Tilde{i}_{(\mu)[\rho\sigma]} = 0 \;
 ,\eqno{(5.14)}
$$
$$
 \nabla_{[\lambda\kappa]} \Tilde{i}_{(\mu)[\rho\sigma]} =
 \partial_{[\lambda\kappa]} \Tilde{i}_{(\mu)[\rho\sigma]} -
 2 \Gamma^{\Hat{\gamma}}_{ \; \cdot \;
 [\lambda\kappa][\rho\sigma]} c_{(\mu)\Hat{\gamma}} -
 \Gamma^{[\tau\omega]}_{ \; \; \cdot \; \;
 [\lambda\kappa][\rho\sigma]} \Tilde{i}_{(\mu)[\tau\omega]} = 0 \;
 ,\eqno{(5.15)}
$$
Conditions (5.5), which also should be fulfilled, are consequence
of (5.9)-(5.15). Connections $\Gamma^{A}_{\; \cdot \; BC}$ are
determined from vanishing of covariant derivatives~\cite{Jan}:
$$
 G_{AB \; ; \; C} = 0 \; , \; \; {{\mathop{g}\limits_{e}}^{A}}_{;
 \; B} = 0 \; , \; \; (A,B = (\mu), \Hat{\alpha}, [\lambda\kappa])
 \; ,\eqno{(5.16)}
$$
where $G_{AB}$ is degenerate metric of original 14-dimensional
space ($\mathrm{rank}(G_{AB}) = 4$), and
${\mathop{g}\limits_{e}}^{A}$ ($e=1,2,...,10$) are ten linearly
independent solutions of the equation $G_{AB}
{\mathop{g}\limits_{e}}^{B} = 0$, which gives
$$
 {\mathop{g}\limits_{e}}^{(\mu)} = - \eta^{(\mu)(\nu)} \left [
 c_{(\nu)\Hat{\alpha}} {\mathop{g}\limits_{e}}^{\Hat{\alpha}} +
 \frac{1}{2} \Tilde{i}_{(\mu)[\lambda\kappa]}
 {\mathop{g}\limits_{e}}^{[\lambda\kappa]} \right ] \;
 ,\eqno{(5.17)}
$$
with the requirement $\mathrm{det}({\mathop{g}\limits_{e}}^{a})
\neq 0$ ($a = \Hat{\alpha}, \; [\lambda\kappa]$). For
$\Gamma^{A}_{\; \cdot \; BC} = \Gamma^{A}_{\; \cdot \; CB}$ we
receive following expressions:
$$
 \Gamma^{(\lambda)}_{\; \cdot \; (\mu)(\nu)} = \frac{1}{2}
 \eta^{(\lambda)(\kappa)} [ \partial_{(\mu)} \eta_{(\kappa)(\nu)}
 + \partial_{(\nu)} \eta_{(\mu)(\kappa)} - \partial_{(\kappa)}
 \eta_{(\mu)(\nu)} ] \; ,\eqno{(5.18)}
$$
$$
 \Gamma^{(\lambda)}_{\; \cdot \; ab} = 0 \; .\eqno{(5.19)}
$$
As it was shown above, imposing of orthogonality conditions (4.6),
(4.14), (4.15) leads to equivalency of metrics{\footnote {Metrics,
which give the same equations of motion, are equivalent. If they
are not degenerate, then equivalence reduces to proportionality of
the metrics with constant coefficient of proportionality.}})
(3.38) (or (3.45)) and (4.47) and gives following expressions for
the rest coefficients of connection:
$$
 \Gamma^{a}_{\; \cdot \; (\mu)(\nu)} = 0 \; , \; \; \Gamma^{a}_{\;
 \cdot \; (\mu)b} = 0 \; , \; \; \Gamma^{[\mu\nu]}_{\; \; \cdot \;
 \; \Hat{\alpha}\Hat{\beta}} = 0 \; ,\eqno{(5.20)}
$$
$$
 \Gamma^{\Hat{\alpha}}_{\; \cdot \; \Hat{\beta}\Hat{\gamma}} =
 \frac{1}{2} h^{\Hat{\alpha}\Hat{\delta}} [ \partial_{\Hat{\beta}}
  h_{\Hat{\delta}\Hat{\gamma}} + \partial_{\Hat{\gamma}}
  h_{\Hat{\beta}\Hat{\delta}} - \partial_{\Hat{\delta}}
  h_{\Hat{\beta}\Hat{\gamma}} ] \doteq H^{\Hat{\alpha}}_{\; \cdot
  \; \Hat{\beta}\Hat{\gamma}} \; ,\eqno{(5.21)}
$$
$$
 \Gamma^{\Hat{\alpha}}_{\; \cdot \; \Hat{\beta}[\lambda\kappa]} =
 \frac{1}{4} h^{\Hat{\alpha}\Hat{\delta}}
 \partial_{[\lambda\kappa]} h_{\Hat{\delta}\Hat{\beta}} \;
 ,\eqno{(5.22)}
$$
$$
 \Gamma^{\Hat{\alpha}}_{\; \cdot \; [\mu\nu][\lambda\kappa]} =
 \frac{1}{8} h^{\Hat{\alpha}\Hat{\delta}} \partial_{\Hat{\delta}}
 \Tilde{\Tilde{j}}_{[\mu\nu][\lambda\kappa]} \; ,\eqno{(5.23)}
$$

\parindent=24pt\ Coefficients of connection $\Gamma^{[\mu\nu]}_{\;
\; \cdot \; \; \Hat{\alpha}[\lambda\kappa]}$ and
$\Gamma^{[\mu\nu]}_{\; \; \cdot \; \;
[\lambda\kappa][\rho\sigma]}$ may be found from the metric (4.47).
Then equivalency of two metrics gives a system of differential
equations for ${\mathop{g}\limits_{e}}^{a}$:
$$
 \partial_{\Hat{\beta}} {\mathop{g}\limits_{e}}^{\Hat{\alpha}} +
 H^{\Hat{\alpha}}_{\; \cdot  \; \Hat{\beta}\Hat{\gamma}}
 {\mathop{g}\limits_{e}}^{\Hat{\gamma}} + \frac{1}{2}
 \Gamma^{\Hat{\alpha}}_{\; \cdot  \; \Hat{\beta}[\lambda\kappa]}
 {\mathop{g}\limits_{e}}^{[\lambda\kappa]} = 0 \; ,\eqno{(5.24)}
$$
$$
 \partial_{[\lambda\kappa]} {\mathop{g}\limits_{e}}^{\Hat{\alpha}}
 + \Gamma^{\Hat{\alpha}}_{\; \cdot  \; [\lambda\kappa]\Hat{\gamma}}
 {\mathop{g}\limits_{e}}^{\Hat{\gamma}} + \frac{1}{2}
 \Gamma^{\Hat{\alpha}}_{\; \cdot \; [\lambda\kappa][\rho\sigma]}
 {\mathop{g}\limits_{e}}^{[\rho\sigma]} = 0 \; ,\eqno{(5.25)}
$$
$$
 \partial_{\Hat{\alpha}} {\mathop{g}\limits_{e}}^{[\lambda\kappa]}
 + \frac{1}{2} \Gamma^{[\lambda\kappa]}_{\; \cdot  \;
 \Hat{\alpha}[\rho\sigma]} {\mathop{g}\limits_{e}}^{[\rho\sigma]}
 = 0 \; ,\eqno{(5.26)}
$$
$$
 \partial_{[\mu\nu]} {\mathop{g}\limits_{e}}^{[\lambda\kappa]} +
 \Gamma^{[\lambda\kappa]}_{\; \; \cdot \; \; [\mu\nu]\Hat{\gamma}}
 {\mathop{g}\limits_{e}}^{\Hat{\gamma}} + \frac{1}{2}
 \Gamma^{[\lambda\kappa]}_{\; \; \cdot \; \; [\mu\nu][\rho\sigma]}
 {\mathop{g}\limits_{e}}^{[\rho\sigma]} = 0 \; .\eqno{(5.27)}
$$

\parindent=24pt\ Equations of motion for the case above may be
written in the form
$$
 \frac{D_{\eta} p_{(\mu)}}{cd\tau} = \frac{dp_{(\mu)}}{cd\tau} -
 \Gamma^{(\rho)}_{\; \cdot \; (\mu)(\nu)} U^{(\nu)} p_{(\rho)} = 0
 \; ,\hspace{32mm}\eqno{(5.28)}
$$
$$
 \frac{ds_{[\mu\nu]}}{cd\tau} = - \frac{\Sigma c}{8}
 \partial_{[\mu\nu]}
 \Tilde{\Tilde{j}}_{[\lambda\kappa][\rho\sigma]}
 \Omega^{[\lambda\kappa]} \Omega^{[\rho\sigma]} + \frac{\Sigma
 c}{2} \partial_{[\mu\nu]} h_{\Hat{\alpha}\Hat{\beta}}
 V^{\Hat{\alpha}} V^{\Hat{\beta}} \; ,\eqno{(5.29)}
$$
$$
 \frac{D_{h} \pi_{\Hat{\alpha}}}{cd\tau} =
 \frac{d\pi_{\Hat{\alpha}}}{cd\tau} - H^{\Hat{\gamma}}_{\; \cdot
 \; \Hat{\alpha}\Hat{\beta}} V^{\Hat{\beta}} \pi_{\Hat{\gamma}} =
  - \frac{\Sigma c}{8} \partial_{\Hat{\alpha}}
 \Tilde{\Tilde{j}}_{[\lambda\kappa][\rho\sigma]}
 \Omega^{[\lambda\kappa]} \Omega^{[\rho\sigma]} \;
 ,\hspace{2mm}\eqno{(5.30)}
$$
where
$$
 p_{(\mu)} = -\frac{\partial L}{c \partial U^{(\mu)}} = \Sigma c
 \eta_{(\mu)(\nu)} U^{(\nu)} = p^{\mathrm{C}}_{(\mu)} \;
 ,\hspace{7mm}\eqno{(5.31)}
$$
$$
 s_{[\lambda\kappa]} = -\frac{\partial L}{c \partial
 \Omega^{[\lambda\kappa]}} = - \frac{\Sigma c}{4}
 \Tilde{\Tilde{j}}_{[\lambda\kappa][\rho\sigma]}
 \Omega^{[\rho\sigma]} = s^{\mathrm{rot}}_{[\lambda\kappa]} \;
 ,\eqno{(5.32)}
$$
$$
 \pi_{\Hat{\alpha}} = -\frac{\partial L}{c \partial
 V^{\Hat{\alpha}}} = \Sigma c h_{\Hat{\alpha}\Hat{\beta}}
 V^{\Hat{\beta}} = \pi^{\mathrm{M}}_{\Hat{\alpha}} \;
 .\hspace{15mm}\eqno{(5.33)}
$$
The requirement of conservation of proper angular momentum of an
extended object gives conditions
$$
 \partial_{[mn]} \Tilde{\Tilde{j}}_{[\lambda\kappa][\rho\sigma]} =
 0 \; , \; \; \partial_{[mn]} h_{\Hat{\alpha}\Hat{\beta}} = 0 \;
 ,\eqno{(5.34)}
$$
$$
 \partial_{[0n]} \left [ h_{\Hat{\alpha}\Hat{\beta}}
 V^{\Hat{\alpha}} V^{\Hat{\beta}} - \frac{1}{4}
 \Tilde{\Tilde{j}}_{[\lambda\kappa][\rho\sigma]}
 \Omega^{[\lambda\kappa]} \Omega^{[\rho\sigma]} \right ] = 0 \;
 .\eqno{(5.35)}
$$

\parindent=24pt\ In the variant considered above, as before, there
takes place geodesic motion of the center of inertia in ${\bf
R}_{4}$. The motion of the point M in $\Hat{{\bf R}}_{4}$, being
not geodesic in general, becomes geodesic one when
$\Omega^{[\mu\nu]} = 0$.

\section{ Internal metric}

\parindent=24pt\ \; \; \; Up to this place a structure and size
of extended objects were bounded by nothing. Therefore in this
Section we shall discuss properties of internal space and try to
build its metric in the case when spatial rotations are absent,
i.e. when $\Omega^{[ab]} = 0$.

\parindent=24pt\ Internal space $\Hat{{\bf R}}_{4}$ is covered
with vectors $r^{\Hat{\alpha}}$ of arbitrary points M relative to
the center of inertia C. In the interval (4.47), describing a
motion of the point M, quantity $\tau$ is proper time of the
center of inertia. Therefore this interval should have the same
value for another point $\mathrm{M}'$ from the manifold $\Hat{{\bf
R}}_{4}$, i.e. for all vectors $r^{\Hat{\alpha}}$ labelling all
points of an extended object. In other words, this interval must
be invariant relative to any admissible transformations in
$\Hat{{\bf R}}_{4}$. Thus, all points of extended object have the
same proper time coinciding with the proper time of the center of
inertia. In Section 4 there were shown that it is possible
provided condition (4.17) be implied. In the case above this
condition, having the form (equation (4.34))
$$
 a^{2} {\bf V}^{4}_{\mathrm{C}} [{\bf r} \times {\bf
 V}_{\mathrm{C}}]^{2} + h_{\Hat{\alpha}\Hat{\beta}}
 V^{\Hat{\alpha}}_{\mathrm{C}} V^{\Hat{\beta}}_{\mathrm{C}} = 0 \;
 ,\eqno{(6.1)}
$$
when $\Omega^{[mn]} = 0$, should be considered along with
orthogonality condition (4.15), reducing to condition (4.30).

\parindent=24pt\ Because a particle is bounded in space-time
${\bf R}_{4}$, one should choose as $\Hat{{\bf R}}_{4}$ manifolds
with finite geodesics (not necessarily closed ones), which are
bounded by some domain in ${\bf R}_{4}$ when the center of inertia
is in rest. Such a domain may be considered as interiority of an
extended particle. Obviously, this domain will seem as world
four-dimensional tube in ${\bf R}_{4}$ in the case of moving
center of inertia. For simplest approximations one may choose the
Minkowski space-time as $\Hat{{\bf R}}_{4}$, corresponding to
infinitely extended particle, internal substance of which is
moving with constant velocity, for $b^{\mu}_{\; \cdot \;
\Hat{\alpha}}$ in this case form a matrix of Lorentz
transformations (see formulae (2.1), (2.9) and (3.35)). Although
such an approximation, certainly, does not both correspond to
reality and satisfy to condition of finiteness of geodesics, it
may lead in quantum case to sufficiently simple relativistic wave
equations of harmonic oscillator type.

\parindent=24pt\ Usually physical bodies and elementary particles
are considered to be extended in three-dimensional space, and this
fact is observable. Theoretically it is attained by imposing of
specific limitations (such as Yukawa condition $p_{\mu} x^{\mu} =
0$, which looks in our case as $\eta^{(\mu)(\nu)} c_{(\nu)\alpha}
p_{\mu} r^{\Hat{\alpha}} = 0$) to relativistic description for
elimination of temporal excitations. In the model under
consideration conditions (4.30) and (6.1) may be treated as
limitations of mentioned type.

\parindent=24pt\ In so far as we give here a relativistic
description, it is reasonably to suppose, that particles (and
maybe also macroscopic bodies) have an extension in temporal axis
too, so that visible three-dimensional objects are cross sections
of four-dimensional ones by the hyperplane $r^{\Hat{0}} = 0$
\footnote {For the first time an idea of four-dimensionally
extended particles was promoted by M.A.Markov~\cite{Mar}.}. This
idea logically follows from successive interpretation of the space
and time as four-dimensional continuum. It should be noted in this
connection, that components $\varphi^{[0n]}$ are not specified by
the Lorentz transformations. It is naturally to require certain
principle of correspondence to be fulfilled between relativistic
and nonrelativistic descriptions. Obviously, passage from
relativity to nonrelativistic descriptions means not only $U \ll c
$, where $\it U$ is a velocity of the center of inertia, but also
$r^{\Hat{0}} = 0$, which is equivalent to Yukawa condition.

\parindent=24pt\ Let us define square of the vector
$r^{\Hat{\alpha}}$ with the help of components
$c_{(\mu)\Hat{\alpha}}$ of fixed 4-hedron in ${\bf R}_{4}$ (see
(3.37)) by means of the formula
$$
 \mathrm{r}^{2} = \eta^{(\mu)(\nu)} c_{(\mu)\Hat{\alpha}}
 c_{(\nu)\Hat{\beta}} r^{\Hat{\alpha}} r^{\Hat{\beta}} =
 h^{0}_{\Hat{\alpha}\Hat{\beta}} r^{\Hat{\alpha}} r^{\Hat{\beta}}
 \; ,\eqno{(6.2)}
$$
where $h^{0}_{\Hat{\alpha}\Hat{\beta}} =
h_{\Hat{\alpha}\Hat{\beta}} (r^{\Hat{\gamma}} = 0)$ is the metric
of the space, which is associated with deformational geodesic
motion of substance in the vicinity of the center of inertia.
(6.2) represents a squared pseudo-distance between points in ${\bf
R}_{4}$ relative to background metric. It reduces to
$\mathrm{r}^{2} = (r^{\Hat{0}})^{2} - {\bf r}^{2}$ in the
Minkowski space. In the non-relativistic limit the quantity
$\mathrm{r}^{2}$ should reduce to $\mathrm{r}^{2}_{\mathrm{NR}} =
h^{0}_{\Hat{a}\Hat{b}} r^{\Hat{a}} r^{\Hat{b}} < 0$, \; $\Hat{a},
\; \Hat{b} = \Hat{1}, \Hat{2}, \Hat{3}$ \;
($\mathrm{r}^{2}_{\mathrm{NR}} = - {\bf r}^{2}$ in the Minkowski
space). Therefore one should require the vector $r^{\Hat{\alpha}}$
to be space-like one, $\mathrm{r}^{2} < 0$ (see also~\cite{Born}).
If $r^{\Hat{\alpha}}$ will be time-like vector, then $r^{\Hat{0}}
= 0$ cannot be assumed in this case, i.e. one cannot pass to
non-relativistic limit. However, one may put $r^{\Hat{a}} = 0$,
which corresponds to collapsing of the object to a point in
three-dimensional space $\Hat{{\bf R}}_{3} \subset \Hat{{\bf
R}}_{4}$. But it means not passage to mass point, but a
disappearance of the object from $\Hat{{\bf R}}_{3}$. Therefore,
for confining of the domain in $\Hat{{\bf R}}_{4} \subset {\bf
R}_{4}$ one should put
$$
 -R^{2}_{0} < \mathrm{r}^{2} = h_{\Hat{\alpha}\Hat{\beta}}
 r^{\Hat{\alpha}} r^{\Hat{\beta}} \leq 0\eqno{(6.3)}
$$
for any vector $r^{\Hat{\alpha}}$ from the manifold $\Hat{{\bf
R}}_{4}$. Here $\mathrm{r}^{2} = 0$ implies $r^{\Hat{\alpha}} =
0$, i.e. motion of the point $\mathrm{M}$ coincides in this case
with the motion of the center of inertia.

\parindent=24pt\ The interval (4.47) must be invariant under any
transformation of the vector $r^{\Hat{\alpha}}$, satisfying to
condition (6.3). Relevant transformation may be written similar to
formula of velocity composition in Special
Relativity~\cite{Tar1},~\cite{Tar2}. To do so we shall consider
five-dimensional space with the interval
$$
 ds^{2} = \eta_{ab} d\xi^{a} d\xi^{b} =
 h_{\Hat{\alpha}\Hat{\beta}} d\xi^{\Hat{\alpha}}
 d\xi^{\Hat{\beta}} + \eta_{\Hat{5}\Hat{5}} (d\xi^{\Hat{5}})^{2} =
 \hspace{15mm}
$$
$$
 = (d\xi^{\Hat{0}})^{2} - (d\xi^{\Hat{1}})^{2} -
 (d\xi^{\Hat{2}})^{2} - (d\xi^{\Hat{3}})^{2} -
 (d\xi^{\Hat{5}})^{2} \; ,\eqno{(6.4)}
$$
where $\eta_{\Hat{5}\Hat{5}} = \pm 1$. In this space one may
consider analogs of inertial reference frame of standard Minkowski
space. Let $\rho^{\Hat{\alpha}}$ denotes a 3+1-dimensional analog
of 3-vector of relative velocity of inertial frames. Like as
coordinates in two reference frames are coupled by the Lorentz
transformation, where 3-vector of relative velocity is a
parameter, in our case an analog of the Lorentz transformation
will have the form~\cite{Tar1}
$$
 {\bf L} = {\bf 1} + \frac{\gamma}{R_{0}} \; \Hat{\mathbf{\rho}} +
 \frac{\gamma -1}{R^{2}_{0} \beta^{2}} \; \Hat{\mathbf{\rho}}^{2}
 \; ,\eqno{(6.5)}
$$
where $\gamma = (1 - \beta^{2})^{-1/2}$, $\Hat{{\bf \rho}} =
h^{0}_{\Hat{\alpha}\Hat{\beta}} \rho^{\Hat{\alpha}} ({\bf
e}^{\Hat{\beta}\Hat{5}} - {\bf e}^{\Hat{5}\Hat{\beta}})$, ${\bf
e}^{ab}$ are elements of complete matrix algebra, satisfying to
relations~\cite{Bog1},~\cite{Bog2}
$$
 {\bf e}^{ab} {\bf e}^{cd} = \eta^{bc} {\bf e}^{ad} \; , \; \;
 ({\bf e}^{ab})^{c}_{\; \cdot \; d} = \eta^{ab} \delta^{b}_{\;
 \cdot \; d}\; , \; \; (a,...,d =
 \Hat{0},\Hat{1},\Hat{2},\Hat{3},\Hat{5}) \; .\eqno{(6.6)}
$$
Quantity
$$
 \beta^{2} = \frac{1}{2 R^{2}_{0}} \mathrm{Sp}({\bf \rho}^{2}) = -
 \frac{\eta^{\Hat{5}\Hat{5}}}{R^{2}_{0}}
 \eta_{\Hat{\alpha}\Hat{\beta}} \rho^{\Hat{\alpha}}
 \rho^{\Hat{\beta}}= - \frac{\eta^{\Hat{5}\Hat{5}}}{R^{2}_{0}}
 h^{0}_{\Hat{\alpha}\Hat{\beta}} \rho^{\Hat{\alpha}}
 \rho^{\Hat{\beta}} = - \frac{\eta^{\Hat{5}\Hat{5}}}{R^{2}_{0}}
 [(\rho^{\Hat{0}})^{2} - {\bf \rho}^{2}] = -
 \frac{\eta^{\Hat{5}\Hat{5}}
 {\mathrm{\rho}}^{2}}{R^{2}_{0}}\eqno{(6.7)}
$$
varies in the interval $0 \leq \beta^{2} < 1$, i.e. $-R^{2}_{0} <
\rho^{2} \leq 0$ \; for $\eta_{\Hat{5}\Hat{5}} = +1$ and $0 \leq
\rho^{2} < R^{2}_{0}$ \; for $\eta_{\Hat{5}\Hat{5}} = -1$.
Obviously, in the case under consideration one should choose
$\eta_{\Hat{5}\Hat{5}} = +1$. Defining in systems $\mathrm{K}$ and
$\mathrm{K}'$ quantities, similar to velocities
$$
 r^{\Hat{\alpha}} = R_{0} \frac{d \xi^{\Hat{\alpha}}}{d
 \xi^{\Hat{5}}} \; , \; \; 'r^{\Hat{\alpha}} = R_{0} \frac{d \;
 '\xi^{\Hat{\alpha}}}{d \; '\xi^{\Hat{5}}} \; ,\eqno{(6.8)}
$$
we find a transformation of the velocity composition law type
$$
 'r^{\Hat{\alpha}} = R_{0} \frac{L^{\Hat{\alpha}}_{\; \cdot \;
 \Hat{\beta}} r^{\Hat{\beta}} + R_{0} L^{\Hat{\alpha}}_{\; \cdot
 \; \Hat{5}}}{L^{\Hat{5}}_{\; \cdot \; \Hat{\beta}}
 r^{\Hat{\beta}} + R_{0} L^{\Hat{5}}_{\; \cdot \; \Hat{5}}} =
 \frac{r^{\Hat{\alpha}} + \rho^{\Hat{\alpha}} + (\gamma -1) \left
 [1 - \frac{h^{0}_{\Hat{\beta}\Hat{\gamma}} r^{\Hat{\beta}}
 \rho^{\Hat{\gamma}}}{\beta^{2} R^{2}_{0}} \right ]
 \rho^{\Hat{\alpha}}}{\gamma \left [1 -
 \frac{h^{0}_{\Hat{\beta}\Hat{\gamma}} r^{\Hat{\beta}}
 \rho^{\Hat{\gamma}}}{R^{2}_{0}} \right ]} \; .\eqno{(6.9)}
$$
This is a desired transformation, which should be completed with
three-dimensional rotations of the vector $r^{\Hat{\alpha}}$. It
is easily to see, that vectors $r^{\Hat{\alpha}}$ and
$'r^{\Hat{\alpha}}$ satisfy condition (6.3). Thus, variations of
parameters $\rho^{\Hat{\alpha}}$ and parameters, corresponding to
three-dimensional rotations of the vectors $r^{\Hat{\alpha}}$,
give all points of the space $\Hat{\bf R}_{4}$.

\parindent=24pt\ Now we require the interval (4.47) to be
invariant under these transformations. Thereby a domain, occupied
by four-dimensionally extended object, is bounded with size
$R_{0}$. Because temporal sizes are not observable $R_{0}$ be
maximal radius of observable three-dimensional cross section of
this object. As it follows from (6.9), if points $\mathrm{M}$ and
$\mathrm{M}'$ are at a short distance, infinitesimal
transformation, connecting their radius vectors $r^{\Hat{\alpha}}$
and $'r^{\Hat{\alpha}}$, has the form
$$
 'r^{\Hat{\alpha}} = r^{\Hat{\alpha}} \left [1 +
 \frac{h^{0}_{\Hat{\beta}\Hat{\gamma}} r^{\Hat{\beta}}
 \rho^{\Hat{\gamma}}}{R^{2}_{0}} \right ] + \rho^{\Hat{\alpha}} \;
 .\eqno{(6.10)}
$$
Variation $d \; 'r^{\Hat{\alpha}}$ in the point $\mathrm{M}'$ may
be found from differential of (6.10) when $\rho^{\Hat{\alpha}} =
\mathrm{const}$:
$$
 d \; 'r^{\Hat{\alpha}} = \left [ \left (1 +
 \frac{h^{0}_{\Hat{\gamma}\Hat{\delta}} r^{\Hat{\gamma}}
 \rho^{\Hat{\delta}}}{R^{2}_{0}} \right )
 \delta^{\Hat{\alpha}}_{\; \cdot \; \Hat{\beta}} +
 \frac{r^{\Hat{\alpha}} h^{0}_{\Hat{\beta}\Hat{\gamma}}
 \rho^{\Hat{\gamma}}}{R^{2}_{0}} \right ] dr^{\Hat{\beta}} \;
 .\eqno{(6.11)}
$$

\parindent=24pt\ Value of a metric $h_{\Hat{\alpha}\Hat{\beta}}$
in the point $\mathrm{M}'$ is defined by
$$
 h_{\Hat{\alpha}\Hat{\beta}}('r^{\Hat{\sigma}}) =
 h_{\Hat{\alpha}\Hat{\beta}}(r^{\Hat{\sigma}}) + \frac{\partial
 h_{\Hat{\alpha}\Hat{\beta}}}{\partial r^{\Hat{\gamma}}} \left [
 \rho^{\Hat{\gamma}} + r^{\Hat{\gamma}}
 \frac{h^{0}_{\Hat{\delta}\Hat{\sigma}} r^{\Hat{\delta}}
 \rho^{\Hat{\sigma}}}{R^{2}_{0}} \right ] = \hspace{57mm}
$$
$$
 = \left [\delta^{\Hat{\mu}}_{\cdot \; \Hat{\alpha}}
 \delta^{\Hat{\nu}}_{\cdot \; \Hat{\beta}} + (H^{\Hat{\mu}}_{\cdot
 \; \Hat{\alpha}\Hat{\gamma}} \delta^{\Hat{\nu}}_{\cdot
 \; \Hat{\beta}} + H^{\Hat{\nu}}_{\cdot \;
 \Hat{\gamma}\Hat{\beta}} \delta^{\Hat{\mu}}_{\cdot \;
 \Hat{\alpha}}) \left (\rho^{\Hat{\gamma}} + r^{\Hat{\gamma}}
 \frac{h^{0}_{\Hat{\delta}\Hat{\sigma}} r^{\Hat{\delta}}
 \rho^{\Hat{\sigma}}}{R^{2}_{0}} \right ) \right ]
 h_{\Hat{\mu}\Hat{\nu}}(r^{\Hat{\alpha}}) \; .\eqno{(6.12)}
$$
When rotations are absent an invariance of the interval (4.47)
under transformations (6.9) reduces to condition
$h_{\Hat{\alpha}\Hat{\beta}}(r^{\Hat{\sigma}}) dr^{\Hat{\alpha}}
dr^{\Hat{\beta}} = h_{\Hat{\alpha}\Hat{\beta}}('r^{\Hat{\sigma}})
d \; 'r^{\Hat{\alpha}} d \; 'r^{\Hat{\beta}}$. From here we get a
relation
$$
 (h_{\Hat{\alpha}\Hat{\beta}} h^{0}_{\Hat{\gamma}\Hat{\sigma}} +
 h_{\Hat{\alpha}\Hat{\sigma}} h^{0}_{\Hat{\gamma}\Hat{\beta}})
 r^{\Hat{\sigma}} + (R^{2}_{0} \; \delta^{\Hat{\nu}}_{\; \cdot \;
 \Hat{\gamma}} + h^{0}_{\Hat{\gamma}\Hat{\sigma}} r^{\Hat{\sigma}}
 r^{\Hat{\nu}}) H^{\Hat{\delta}}_{\; \cdot \;
 \Hat{\alpha}\Hat{\nu}} h_{\Hat{\delta}\Hat{\beta}} = 0 \;
 ,\eqno{(6.13)}
$$
being actually an equation for metric. Assuming, that the most
general form of the metric is
$$
 h_{\Hat{\alpha}\Hat{\beta}} = \chi_{\Hat{\alpha}\Hat{\beta}} +
 (a_{\Hat{\alpha}} h^{0}_{\Hat{\beta}\Hat{\gamma}} +
 a_{\Hat{\beta}} h^{0}_{\Hat{\alpha}\Hat{\gamma}})
 r^{\Hat{\gamma}} + b \; h^{0}_{\Hat{\alpha}\Hat{\gamma}}
 h^{0}_{\Hat{\beta}\Hat{\delta}} r^{\Hat{\gamma}} r^{\Hat{\delta}}
 \; ,\eqno{(6.14)}
$$
where $\chi_{\Hat{\alpha}\Hat{\beta}}$, $a_{\Hat{\alpha}}$, $\it
b$ are functions only on variables ${\mathrm{r}}^{2} =
h^{0}_{\Hat{\alpha}\Hat{\beta}} r^{\Hat{\alpha}} r^{\Hat{\beta}}$
and $\eta = k_{\Hat{\alpha}} r^{\Hat{\alpha}}$ ($k_{\Hat{\alpha}}$
is constant vector), we find $\partial h_{\Hat{\alpha}\Hat{\beta}}
/ \partial \eta = 0$, $a_{\Hat{\alpha}} = 0$ and
$$
 h_{\Hat{\alpha}\Hat{\beta}} = \left (1 +
 \frac{{\mathrm{r}}^{2}}{R^{2}_{0}} \right )^{-1}
 h^{0}_{\Hat{\alpha}\Hat{\beta}} - \left (1 +
 \frac{{\mathrm{r}}^{2}}{R^{2}_{0}} \right )^{-2} \;
 \frac{h^{0}_{\Hat{\alpha}\Hat{\gamma}}
 h^{0}_{\Hat{\beta}\Hat{\delta}} r^{\Hat{\gamma}}
 r^{\Hat{\delta}}}{R^{2}_{0}} \; ,\eqno{(6.15)}
$$
$$
 h^{\Hat{\alpha}\Hat{\beta}} = \left (1 +
 \frac{{\mathrm{r}}^{2}}{R^{2}_{0}} \right ) \left [
 h_{0}^{\Hat{\alpha}\Hat{\beta}} + \frac{r^{\Hat{\alpha}}
 r^{\Hat{\beta}}}{R^{2}_{0}} \right ] \; , \; \; \;
 h_{\Hat{\alpha}\Hat{\gamma}} h^{\Hat{\gamma}\Hat{\beta}} =
 \delta^{\Hat{\beta}}_{ \; \cdot \; \Hat{\alpha}} \; .\eqno{(6.16)}
$$

\parindent=24pt\ Let us find field equation for this metric.
Connections, curvature tensor, Ricci tensor, scalar curvature and
Einstein tensor are
$$
 H^{\Hat{\alpha}}_{ \; \cdot \; \Hat{\beta}\Hat{\gamma}} =
 -\frac{1}{R^{2}_{0}} \left (1 +
 \frac{{\mathrm{r}}^{2}}{R^{2}_{0}} \right )^{-1}
 (\delta^{\Hat{\alpha}}_{ \; \cdot \; \Hat{\beta}}
 h^{0}_{\Hat{\gamma}\Hat{\delta}} + \delta^{\Hat{\alpha}}_{
 \; \cdot \; \Hat{\gamma}} h^{0}_{\Hat{\beta}\Hat{\delta}})
 r^{\Hat{\delta}} \; .\eqno{(6.17)}
$$
$$
 R^{\Hat{\mu}}_{ \; \cdot \; \Hat{\nu}\Hat{\alpha}\Hat{\beta}} =
 \partial_{[\Hat{\alpha}} H^{\Hat{\mu}}_{ \; \cdot \;
 |\Hat{\nu}|\Hat{\beta}]} + H^{\Hat{\mu}}_{ \; \cdot \;
 \Hat{\gamma}[\Hat{\alpha}} H^{\Hat{\gamma}}_{ \; \cdot \;
 |\Hat{\nu}|\Hat{\beta}]} = \frac{\eta}{R^{2}_{0}}
 (h_{\Hat{\nu}\Hat{\beta}} \delta^{\Hat{\mu}}_{ \; \cdot \;
 \Hat{\alpha}} - h_{\Hat{\nu}\Hat{\alpha}} \delta^{\Hat{\mu}}_{ \;
 \cdot \; \Hat{\beta}}) \; ,\eqno{(6.18)}
$$
$$
 R_{\Hat{\alpha}\Hat{\beta}} = R^{\Hat{\mu}}_{ \; \cdot \;
 \Hat{\alpha}\Hat{\mu}\Hat{\beta}} = \frac{3}{R^{2}_{0}}
 h_{\Hat{\alpha}\Hat{\beta}} = \lambda h_{\Hat{\alpha}\Hat{\beta}}
 \; ,\eqno{(6.19)}
$$
$$
 R = h^{\Hat{\alpha}\Hat{\beta}} R_{\Hat{\alpha}\Hat{\beta}} =
 \frac{12}{R^{2}_{0}} \; ,\eqno{(6.20)}
$$
$$
 G_{\Hat{\alpha}\Hat{\beta}} = R_{\Hat{\alpha}\Hat{\beta}} -
 \frac{1}{2} R h_{\Hat{\alpha}\Hat{\beta}} = -\frac{3}{R^{2}_{0}}
 h_{\Hat{\alpha}\Hat{\beta}} = - \kappa_{0}
 T_{\Hat{\alpha}\Hat{\beta}} \; ,\eqno{(6.21)}
$$
respectively, where $T_{\Hat{\alpha}\Hat{\beta}} = 3
\kappa^{-1}_{0} R^{2}_{0} h_{\Hat{\alpha}\Hat{\beta}} = \lambda
\kappa^{-1}_{0} h_{\Hat{\alpha}\Hat{\beta}}$ is the
energy-momentum tensor, $\lambda = 3 R^{2}_{0}$ is cosmological
constant, $\kappa_{0} = 8 \pi G / c^{4} = 2,0759 \cdot 10^{-43} \;
\mathrm{N}^{-1}$ is Einstein gravitational constant. Equation
(6.19) shows, that the Ricci tensor is of $\lambda$-term type,
while the energy-momentum tensor describes a perfect fluid with
pressure $p = \lambda \kappa^{-1}_{0}$ and energy density $\mu$,
satisfying to relation $p + \mu = 0$ (\cite{Kra}, § 5.2), which
gives $\mu = -\lambda \kappa^{-1}_{0} < 0$. To satisfy a condition
of positiveness of the energy density, we shall make a
substitution $\mu' = \mu + \Lambda \kappa^{-1}_{0}$, $p' = p -
\Lambda \kappa^{-1}_{0}$, so that condition $\mu' > 0$ be
fulfilled. Then, obviously, $p' + \mu' = p + \mu = 0$ and equation
(6.21) will be written down in the next form
$$
 R_{\Hat{\alpha}\Hat{\beta}} - \frac{1}{2} R
 h_{\Hat{\alpha}\Hat{\beta}} + \Lambda h_{\Hat{\alpha}\Hat{\beta}}
 = - \left (\frac{3}{R^{2}_{0}} - \Lambda \right )
 h_{\Hat{\alpha}\Hat{\beta}} = - \kappa_{0}
 T'_{\Hat{\alpha}\Hat{\beta}} = - \kappa_{0}
 T_{\Hat{\alpha}\Hat{\beta}} + \Lambda h_{\Hat{\alpha}\Hat{\beta}}
 \; .\eqno{(6.22)}
$$
Thus, introduction of cosmological constant is rather artificial
step here.

\parindent=24pt\ We shall say also some words about the
requirement $\mu > 0$. It is stipulated by the fact, that Einstein
equations (6.21) describe, as it is generally adopted, an external
field of gravitating masses, obeying to the Newton's world gravity
law ${\bf F}_{\gamma} = -G m_{1} m_{2} r^{-3} {\bf r}$ in the
non-relativistic approximation. However, Einstein equations have
wider interpretation, for they describe arbitrary curved space
(without torsion). Particularly, one may consider they to describe
also a field of electric charge, which in the non-relativistic
limit of point charge reduces to the Coulomb law ${\bf F}_{e} = (4
\pi \varepsilon_{0} r^{3})^{-1} e_{1} e_{2} {\bf r}$. Taking into
account that Newton's and Coulomb's laws have the same form and
differ, besides of the sense of quantities, with only sign, it is
reasonably to assume, that in the case, when Einstein equations
describe a field of electron, the constant $\it G$ should be
substituted with $-e^{2} (4 \pi \varepsilon_{0} m^{2}_{e})^{-1} =
-\alpha \hbar c / m^{2}_{e} \approx -2,78 \cdot 10^{32} \;
\mathrm{m}^{3}/(\mathrm{kg} \cdot \mathrm{s}^{2})$, where $m_{e}$
is the electron rest mass. Then the constant $\kappa_{0}$ will
have the next value
$$
 \kappa_{0} = - \frac{8 \pi e^{2}}{4 \pi \varepsilon_{0} m^{2}_{e}
 c^{4}} = - \frac{4 \alpha h}{m^{2}_{e} c^{3}} \approx -0,865 \;
 \mathrm{N}^{-1} \; .\eqno{(6.23)}
$$

\parindent=24pt\ Since energy density is the
$\Hat{0}\Hat{0}$-component of the energy-momentum tensor
$$
 \mu = T^{\Hat{0}\Hat{0}} = h^{\Hat{0}\Hat{\alpha}}
 h^{\Hat{0}\Hat{\beta}} T_{\Hat{\alpha}\Hat{\beta}} =
 -\frac{3}{\kappa_{0} R^{2}_{0}} \left (1 +
 \frac{(r^{\Hat{0}})^{2} - {\bf r}^{2}}{R^{2}_{0}} \right ) \left
 (1 + \frac{(r^{\Hat{0}})^{2}}{R^{2}_{0}} \right ) \;
 ,\eqno{(6.24)}
$$
then mass of extended object will be equal to
$$
 m_{0} = \frac{1}{c^{2}} {\mathop{\int}\limits_{V}} \mu dV =
 -\frac{3}{\kappa_{0} c^{2} R^{2}_{0}} {\mathop{\int}\limits_{V}}
 \left (1 + \frac{{\mathrm{r}}^{2}}{R^{2}_{0}} \right ) \left (1 +
 \frac{(r^{\Hat{0}})^{2}}{R^{2}_{0}} \right ) dV \; .\eqno{(6.25)}
$$
If $\mu$ does not depend on $r^{\Hat{0}}$ (what is rather strong
condition), then (6.24) leads to
$$
 \frac{\partial}{\partial r^{\Hat{0}}} \left [ \left (1 +
 \frac{(r^{\Hat{0}})^{2} - {\bf r}^{2}}{R^{2}_{0}} \right ) \left
 (1 + \frac{(r^{\Hat{0}})^{2}}{R^{2}_{0}} \right ) \right ] =
 \frac{2 r^{\Hat{0}}}{R^{2}_{0}} \left (2 +
 \frac{2(r^{\Hat{0}})^{2} - {\bf r}^{2}}{R^{2}_{0}} \right ) = 0
 \; ,\eqno{(6.26)}
$$
whence there follow two possibilities
$$
 a) \; r^{\Hat{0}} = 0 \; ,\eqno{(6.27)}
$$
and
$$
 b) \; 1 + \frac{(r^{\Hat{0}})^{2}}{R^{2}_{0}} =
 \frac{{\bf r}^{2}}{2R^{2}_{0}} \; .\eqno{(6.28)}
$$

\parindent=24pt\ Condition (6.28) contradicts to condition (6.3),
for combination of them gives
$$
 0 \leq {\bf r}^{2} - (r^{\Hat{0}})^{2} = \frac{{\bf r}^{2}}{2} +
 R^{2}_{0} < R^{2}_{0} \; ,\eqno{(6.29)}
$$
whence it follows, that (6.29) is inconsistent when ${\bf r}^{2}
\neq 0$, while ${\bf r}^{2} = 0$ is inconsistent with (6.28).
Condition (6.27) is consistent with (6.3), which gives $0 \leq
{\bf r}^{2} < R^{2}_{0}$.

\parindent=24pt\ Thus, substitution (6.27) in (6.25) and taking
into account spherical symmetry of the solution leads to
$$
 m_{0} = -\frac{3}{\kappa_{0} c^{2} R^{2}_{0}}
 {\mathop{\int}\limits_{V}} \left (1 - \frac{{\bf
 r}^{2}}{R^{2}_{0}} \right ) dV = \hspace{56mm}
$$
$$
 = -\frac{12 \pi}{\kappa_{0} c^{2} R^{2}_{0}}
 {\mathop{\int}\limits^{R_{0}}_{0}} \left (1 - \frac{{\bf
 r}^{2}}{R^{2}_{0}} \right ) r^{2} dr = -\frac{8 \pi R_{0}}{5
 \kappa_{0} c^{2}} = \frac{R_{0} m^{2}_{e} c}{5 \alpha \hbar} =
 \frac{R_{0} m_{e}}{5 a_{0}} \; ,\eqno{(6.30)}
$$
where $a_{0} = \alpha \hbar / m_{e} c = 2,81751 \cdot 10^{-15} \;
\mathrm{m}$ is the classical radius of the electron. Under
assumption $m_{0} = m_{e}$ we get a following estimation for
radius of an extended object $R_{0} = 5 a_{0} = 1,41 \cdot
10^{-14} \; \mathrm{m}$.

\parindent=24pt\ Calculations above should not attach absolute
importance to, for there were made too many assumptions of
intuitive character such, for example, as independence of the
energy density $\mu$ on $r^{\Hat{0}}$, description of internal
space of the electron by the metric (6.15), interpolation of the
Coulomb law into electron interiority, and so on. It is sufficient
to point out, that modern indirect measurings of the electron
radius give the value $r_{e} = 0,452 \cdot 10^{-22} \;
\mathrm{m}$~\cite{Deh1}-\cite{Deh2}.

\parindent=24pt\ Let us consider now restrictions (4.30) and (6.1)
imposed to velocity $V^{\Hat{\alpha}}_{\mathrm{C}}$ for obtained
internal metric (6.15). In the center-of-inertia system the metric
$g_{\mu\nu}$, describing a motion of the point M in the background
space, must coincide with internal metric, so that $g_{\mu\nu} =
\delta^{\Hat{\alpha}}_{\mu} \delta^{\Hat{\beta}}_{\nu}
h_{\Hat{\alpha}\Hat{\beta}}$, where $\delta^{\Hat{\alpha}}_{\mu} =
1$, if $\mu = \alpha$ and $\delta^{\Hat{\alpha}}_{\mu} = 0$, if
$\mu \neq \alpha$. Then equations (2.13)-(2.14) shall have the
form
$$
  h_{\Hat{\alpha}\Hat{\beta}} e^{\Hat{\alpha}}_{ \; \cdot \;
  (\lambda)} e^{\Hat{\beta}}_{ \; \cdot \; (\kappa)} =
  \eta_{(\lambda)(\kappa)} \; ,\eqno{(6.31)}
$$
$$
  \eta^{(\lambda)(\kappa)} e^{\Hat{\alpha}}_{ \; \cdot \;
  (\lambda)} e^{\Hat{\beta}}_{ \; \cdot \; (\kappa)} =
  h^{\Hat{\alpha}\Hat{\beta}} \; .\eqno{(6.32)}
$$
It is not difficult to show these equations to give following
expressions for components of reciprocal 4-hedron
$\underline{e}^{\mu} = \{ e^{\mu}_{ \; \cdot \; (\lambda)} \} = \{
\delta^{\mu}_{ \; \cdot \; \Hat{\alpha}} e^{\Hat{\alpha}}_{ \;
\cdot \; (\lambda)} \}$:
$$
  e^{\Hat{\alpha}}_{ \; \cdot \; (\mu)} = \delta^{\Hat{\alpha}}_{
  \; \cdot \; \lambda} e^{\lambda}_{ \; \cdot \; (\mu)} = \left (1
  + \frac{\mathrm{r}^{2}}{R^{2}_{0}} \right )^{1/2} \left
  [\delta^{\Hat{\alpha}}_{ \; \cdot \; (\mu)} - \left [1 \pm \left
  (1 + \frac{\mathrm{r}^{2}}{R^{2}_{0}} \right )^{1/2} \right ]
  \frac{\eta_{(\mu)\Hat{\beta}} r^{\Hat{\alpha}}
  r^{\Hat{\beta}}}{\mathrm{r}^{2}} \right ] \; .\eqno{(6.33)}
$$
For components of moving 4-hedron $\overline{e}_{\lambda} = \{
e^{(\mu)}_{ \; \cdot \; \lambda} \} = \{ e^{(\mu)}_{ \; \cdot \;
\Hat{\alpha}} \delta^{\Hat{\alpha}}_{ \; \cdot \; \lambda} \}$,
satisfying to relations
$$
  e^{(\mu)}_{ \; \cdot \; \Hat{\alpha}} e^{\Hat{\alpha}}_{
  \; \cdot \; (\nu)} = \delta^{(\mu)}_{ \; \cdot \; (\nu)} \; , \;
  \; \; e^{\Hat{\alpha}}_{ \; \cdot \; (\mu)} e^{(\mu)}_{ \; \cdot
  \; \Hat{\beta}} = \delta^{\Hat{\alpha}}_{ \; \cdot \;
  \Hat{\beta}} \; ,\eqno{(6.34)}
$$
$$
  \eta_{(\mu)(\nu)} e^{(\mu)}_{ \; \cdot \; \Hat{\alpha}}
  e^{(\nu)}_{ \; \cdot \; \Hat{\beta}} =
  h_{\Hat{\alpha}\Hat{\beta}} \; , \; \; \;
  h^{\Hat{\alpha}\Hat{\beta}} e^{(\mu)}_{ \; \cdot \;
  \Hat{\alpha}} e^{(\nu)}_{ \; \cdot \; \Hat{\beta}} =
  \eta^{(\mu)(\nu)} \; ,\eqno{(6.35)}
$$
we get
$$
  e^{(\mu)}_{ \; \cdot \; \Hat{\alpha}} = \delta^{\beta}_{ \;
  \cdot \; \Hat{\alpha}} e^{(\mu)}_{ \; \cdot \; \beta} = \left (1
  + \frac{\mathrm{r}^{2}}{R^{2}_{0}} \right )^{-1/2} \left
  [\delta^{(\mu)}_{ \; \cdot \; \Hat{\alpha}} - \left [1 \pm \left
  (1 + \frac{\mathrm{r}^{2}}{R^{2}_{0}} \right )^{-1/2} \right ]
  \frac{\delta^{(\mu)}_{ \; \cdot \; \Hat{\beta}}
  h^{0}_{\Hat{\alpha}\Hat{\gamma}} r^{\Hat{\beta}}
  r^{\Hat{\gamma}}}{\mathrm{r}^{2}} \right ] \; .\eqno{(6.36)}
$$

\parindent=24pt\ Substituting (6.15) and (6.33) into (4.30) we
come to the next relation between $V^{\Hat{\alpha}}_{\mathrm{C}}$
and $r^{\Hat{\alpha}}$
$$
  h_{\Hat{\alpha}\Hat{\beta}} e^{\alpha}_{(0)}
  V^{\Hat{\beta}}_{\mathrm{C}} = \left (1 +
  \frac{\mathrm{r}^{2}}{R^{2}_{0}} \right )^{-1/2} \left
  [V^{\Hat{0}}_{\mathrm{C}} - \left [ 1 \pm \left
  (1 + \frac{\mathrm{r}^{2}}{R^{2}_{0}} \right )^{-1/2} \right ]
  \frac{r^{\Hat{0}} h^{0}_{\Hat{\alpha}\Hat{\beta}}
  r^{\Hat{\alpha}} V^{\Hat{\beta}}_{\mathrm{C}}}{\mathrm{r}^{2}}
  \right ] = 0 \; ,\eqno{(6.37)}
$$
which may be written down in the form
$$
  \left [1 - \frac{{\bf r}^{2}}{R^{2}_{0}} \mp \left (1 +
  \frac{\mathrm{r}^{2}}{R^{2}_{0}} \right )^{1/2} \right ]
  V^{\Hat{0}}_{\mathrm{C}} = \frac{r^{\Hat{0}}
  h^{0}_{\Hat{a}\Hat{b}} r^{\Hat{a}}
  V^{\Hat{b}}_{\mathrm{C}}}{R^{2}_{0}} \; ,\eqno{(6.38)}
$$
where $\mathbf{r}^{2} = -h^{0}_{\Hat{a}\Hat{b}} r^{\Hat{a}}
r^{\Hat{b}}$ is determined in (4.22).

\parindent=24pt\ In view of (6.37) squared vector (4.31) is
$$
  \mathbf{V}^{2}_{\mathrm{C}} = -\eta_{(m)(n)} e^{(m)}_{\; \cdot \;
  \Hat{\alpha}} e^{(n)}_{\; \cdot \; \Hat{\beta}}
  V^{\Hat{\alpha}}_{\mathrm{C}} V^{\Hat{\beta}}_{\mathrm{C}} =
  -h_{\Hat{\alpha}\Hat{\beta}} V^{\Hat{\alpha}}_{\mathrm{C}}
  V^{\Hat{\beta}}_{\mathrm{C}} =
$$
$$
  = \left (1 + \frac{\mathrm{r}^{2}}{R^{2}_{0}} \right )^{-1}
  \left \{-h^{0}_{\Hat{a}\Hat{b}} V^{\Hat{a}}_{\mathrm{C}}
  V^{\Hat{b}}_{\mathrm{C}} - \frac{\mathbf{r}^{2}
  (V^{\Hat{0}}_{\mathrm{C}})^{2}}{(r^{\Hat{0}})^{2}} + 2 \left [1
  \mp \left (1 + \frac{\mathrm{r}^{2}}{R^{2}_{0}} \right )^{1/2}
  \right ] \frac{R^{2}_{0}
  (V^{\Hat{0}}_{\mathrm{C}})^{2}}{(r^{\Hat{0}})^{2}} \right \} \;
  ,\eqno{(6.39)}
$$
where only upper sign in square brackets should be left, for lower
sign is inconsistent with obvious inequality ${\bf
V}^{2}_{\mathrm{C}} = -h_{\Hat{\alpha}\Hat{\beta}}
V^{\Hat{\alpha}}_{\mathrm{C}} V^{\Hat{\beta}}_{\mathrm{C}} \geq
0$, where equality sign takes place only when
$V^{\Hat{\alpha}}_{\mathrm{C}} = 0$. Then restriction (6.1) will
be written down as
$$
  (a^{2} {\bf V}^{2}_{\mathrm{C}} [{\bf r} \times {\bf
  V}_{\mathrm{C}}]^{2} -1) h_{\Hat{\alpha}\Hat{\beta}}
  V^{\Hat{\alpha}}_{\mathrm{C}} V^{\Hat{\beta}}_{\mathrm{C}} = 0
  \; ,\eqno{(6.40)}
$$
whence it follows
$$
  a {\bf V}^{2}_{\mathrm{C}} = \pm \sqrt{\frac{{\bf
  V}^{2}_{\mathrm{C}}}{[{\bf r} \times {\bf V}_{\mathrm{C}}]^{2}}}
  = \pm \frac{1}{r \sin \vartheta} \; ,\eqno{(6.41)}
$$
where $\vartheta$ is an angle between vectors $\bf r$ and ${\bf
V}_{\mathrm{C}}$,
$$
  ({\bf r} \cdot {\bf V}_{\mathrm{C}}) = -\eta_{(m)\Hat{a}}
  e^{(m)}_{ \; \cdot \; \Hat{\beta}} r^{\Hat{a}}
  V^{\Hat{\beta}}_{\mathrm{C}} = - \left (1 +
  \frac{\mathrm{r}^{2}}{R^{2}_{0}} \right )^{-1/2} \left [1 -
  \left (1 + \frac{\mathrm{r}^{2}}{R^{2}_{0}} \right )^{-1/2}
  \right ] \frac{R^{2}_{0} V^{\Hat{0}}_{\mathrm{C}}}{r^{\Hat{0}}}
  \; ,\eqno{(6.42)}
$$
$$
  [{\bf r} \times {\bf V}_{\mathrm{C}}]^{2} = {\bf r}^{2} {\bf
  V}^{2}_{\mathrm{C}} - ({\bf r} \cdot {\bf V}_{\mathrm{C}})^{2}
  = \hspace{70mm}
$$
$$
  = \left (1 + \frac{\mathrm{r}^{2}}{R^{2}_{0}} \right )^{-1}
  \left \{-{\bf r}^{2} h^{0}_{\Hat{a}\Hat{b}}
  V^{\Hat{a}}_{\mathrm{C}} V^{\Hat{b}}_{\mathrm{C}} - R^{2}_{0}
  ( V^{\Hat{0}}_{\mathrm{C}} )^{2} + \frac{{\bf r}^{2}( R^{2}_{0}
  - {\bf r}^{2} ) ( V^{\Hat{0}}_{\mathrm{C}} )^2}{{r^{\Hat{0}}}^{2}}
  - \right. \hspace{-17mm}
$$
$$
  - 2 \left [1 - \left (1 + \frac{\mathrm{r}^{2}}{R^{2}_{0}} \right
  )^{1/2} \right ] \left. \frac{R^{2}_{0} ( R^{2}_{0} - {\bf r}^{2} )
  ( V^{\Hat{0}}_{\mathrm{C}} )^{2}}{{r^{\Hat{0}}}^{2}} \right \} \;
  .\hspace{5mm}\eqno{(6.43)}
$$

\parindent=24pt\ Due to $V^{\Hat{\alpha}}_{\mathrm{C}} =
dr^{\Hat{\alpha}} / c d\tau_{\mathrm{C}}$ equation (6.38)
represents in fact an equation of hypersurface, along which a
point $\mathrm{M}$ is moving. Setting $u = r^{\Hat{0}} / R_{0}$,
$A = 1 - {{\bf r}^{2}} / R^{2}_{0}$ we get from (6.38) the next
equation
$$
  2 ( A - \sqrt{A + u^{2}} ) du = u dA \; ,\eqno{(6.44)}
$$
solution of which is
$$
  \sqrt{A + u^{2}} - 1 = g u \; ,\eqno{(6.45)}
$$
or
$$
  (1 - g^{2}) \left [ r^{\Hat{0}} - \frac{g R_{0}}{1 - g^{2}}
  \right]^{2} - {\bf r}^{2} = \frac{g^{2} {R_{0}}^{2}}{1 - g^{2}}
  \; ,\eqno{(6.46)}
$$
where $\it g$ is constant parameterizing the hypersurface.
According to (6.3) $-1 < \sqrt{A + u^{2}} - 1 \leq 0$. Hence $-1 <
gu \leq 0$, or $-R_{0} < g r^{\Hat{0}} \leq 0$. Therefore (6.46)
gives
$$
  0 \leq \frac{r^{\Hat{0}}}{R_{0}} = -\frac{g}{g^{2}-1} -
  \sqrt{\frac{g^{2}}{(g^{2}-1)^{2}} - \frac{{\bf
  r}^{2}}{{R^{2}_{0}} (g^{2}-1)}} \; \leq -\frac{1}{g} \; , \; \;
  g < 0 \; ,\eqno{(6.47)}
$$
$$
  -\frac{1}{g} \leq \frac{r^{\Hat{0}}}{R_{0}} = -\frac{g}{g^{2}-1}
  + \sqrt{\frac{g^{2}}{(g^{2}-1)^{2}} - \frac{{\bf
  r}^{2}}{{R^{2}_{0}} (g^{2}-1)}} \; \leq 0 \; , \; \; g > 0 \;
  .\eqno{(6.48)}
$$
Here $g = \pm \infty$ corresponds to $r^{\Hat{0}} = 0$, and $g =
0$ corresponds to light cone
$$
  (r^{\Hat{0}})^{2} - {\bf r}^{2} = 0 \; .\eqno{(6.49)}
$$
At $g = \pm 1$ equation (6.46) is degenerated into the equation of
paraboloid
$$
  r^{\Hat{0}} = \pm {{\bf r}^{2}}/2R_{0} \; .\eqno{(6.50)}
$$
In all cases internal coordinates vary in the next boundaries
$$
  -\frac{R_{0} \sqrt{g^{2} + 1}}{g} \leq r^{\Hat{a}} \leq
  \frac{R_{0} \sqrt{g^{2} + 1}}{g} \; , \; \; g > 0 \;
  ;\hspace{6mm}
$$
$$
  \frac{R_{0} \sqrt{g^{2} + 1}}{g} \leq r^{\Hat{a}} \leq -
  \frac{R_{0} \sqrt{g^{2} + 1}}{g} \; , \; \; g < 0 \;
  .\eqno{(6.51)}
$$
These hypersurfaces are represented in Fig.1.

\begin{center}
\includegraphics[scale=0.8]{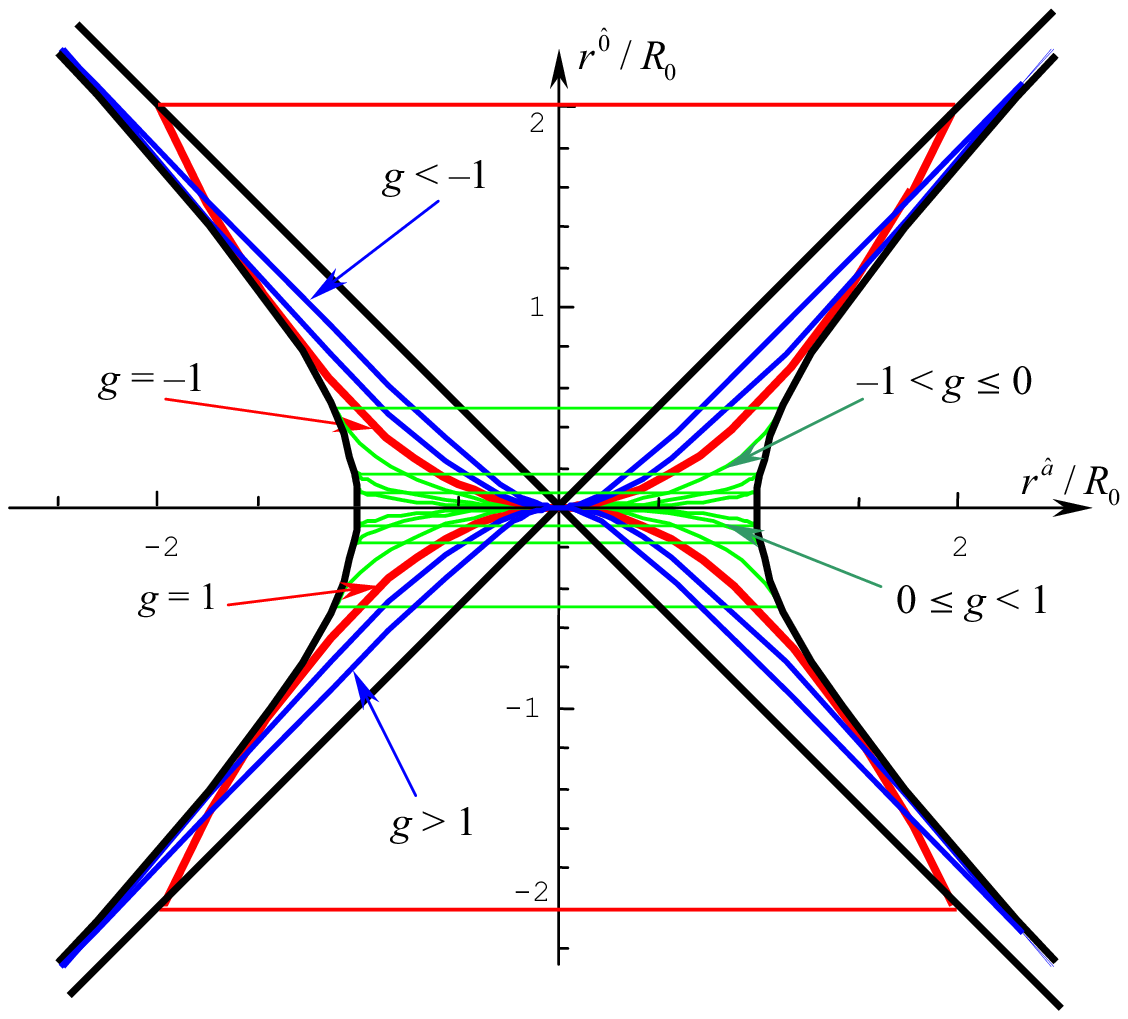} \vspace{-10mm}
\end{center}
\centerline{ Figure 1.} \vspace{10mm}

\parindent=24pt\ Calculations above give a possibility to set a
sense of components $B^{(k)} = \Omega^{[0k]}_{\mathrm{C}}$ forming
the vector $\bf B$. The sign "$\pm$" in (6.41) corresponds both to
the sign of $r^{\Hat{0}}$ and to direction of the pseudo-vector
$[{\bf r} \times {\bf V}_{\mathrm{C}}]$, which is a moment of the
vector ${\bf V}_{\mathrm{C}}$ relative to the center of inertia.
Consequently, substituting (6.41) in (4.33) one may put (when
${\bf \Omega} = {\bf 0}$)
$$
  B^{(k)} = \Omega^{[0k]}_{\mathrm{C}} = |r^{\Hat{0}}|^{-1}
  \sqrt{\frac{{\bf V}^{2}_{\mathrm{C}}}{[{\bf r} \times {\bf
  V}_{\mathrm{C}}]^{2}}} \eta^{(k)(l)} [{\bf r} \times {\bf
  V}_{\mathrm{C}}]_{(l)} = \hspace{35mm}
$$
$$
  = |r^{\Hat{0}}|^{-1} \sqrt{\frac{{\bf
  V}^{2}_{\mathrm{C}}}{[{\bf r} \times {\bf V}_{\mathrm{C}}]^{2}}}
  \eta^{(k)(l)} \varepsilon_{(l)(m)(n)} \delta^{(m)}_{ \; \; \cdot
  \; \; \Hat{a}} r^{\Hat{a}} V^{(n)}_{\mathrm{C}} \;
  ,\eqno{(6.52)}
$$
where $r^{\Hat{0}}$ is determined in (6.47)-(6.50).

\parindent=24pt\ At last, in the case under consideration, when
$j_{[\mu\nu][\lambda\kappa]} = j^{0}_{[\mu\nu][\lambda\kappa]}$
and ${\bf \Omega} = {\bf 0}$, tensor (5.32) has the form
$$
  s^{\mathrm{rot}}_{[\mu\nu]} = -\frac{\Sigma c}{4}
  \Tilde{\Tilde{j}}^{0}_{[\mu\nu][\lambda\kappa]}
  \Omega^{[\lambda\kappa]}_{\mathrm{C}} = -\frac{\Sigma c}{2}
  \left [\eta_{(\mu)\Hat{\alpha}} \eta_{(\nu)(\kappa)} -
  \eta_{(\nu)\Hat{\alpha}} \eta_{(\mu)(\kappa)} \right ]
  \eta_{(\lambda)\Hat{\beta}} r^{\Hat{\alpha}} r^{\Hat{\beta}}
  \Omega^{[\lambda\kappa]}_{\mathrm{C}} = \hspace{30mm}
$$
$$
  = \frac{\Sigma c}{2} \left [ \eta_{(\mu)\Hat{\alpha}}
  (\eta_{(\nu)(0)} \eta_{(k)\Hat{b}} r^{\Hat{b}} - \eta_{(\nu)(k)}
  r^{\Hat{0}}) - \eta_{(\nu)\Hat{\alpha}} (\eta_{(\mu)(0)}
  \eta_{(k)\Hat{b}} r^{\Hat{b}} - \eta_{(\mu)(k)} r^{\Hat{0}})
  \right ] r^{\Hat{\alpha}} \Omega^{[0k]}_{\mathrm{C}} \;
  ,\eqno{(6.53)}
$$
whence it follows
$$
  s^{\mathrm{rot}}_{[mn]} = \frac{\Sigma c}{2} \left [
  \eta_{(n)\Hat{a}} \eta_{(m)(k)} - \eta_{(m)\Hat{a}}
  \eta_{(n)(k)} \right ] r^{\Hat{a}} r^{\Hat{0}}
  \Omega^{[0k]}_{\mathrm{C}} = \varepsilon_{(m)(n)(k)} s^{(k)} \;
  ,\eqno{(6.54)}
$$
$$
  s^{(k)} = \frac{1}{2} \varepsilon^{(m)(n)(k)}
  s^{\mathrm{rot}}_{[mn]} = \frac{\Sigma c}{2} [r^{\Hat{0}} {\bf
  B} \times {\bf r}]^{(k)} = \hspace{34mm}
$$
$$
  = -\frac{\Sigma c \; \mathrm{sign}(r^{\Hat{0}})}{2}
  \sqrt{\frac{{\bf V}^{2}_{\mathrm{C}}}{[{\bf r} \times {\bf
  V}_{\mathrm{C}}]^{2}}} \; [{\bf r} \times [{\bf r} \times {\bf
  V}_{\mathrm{C}}]]^{(k)} \; ,\hspace{14mm}\eqno{(6.55)}
$$
$$
  s^{\mathrm{rot}}_{[0n]} = -\frac{\Sigma c}{2} (r^{\Hat{0}})^{2}
  \eta_{(n)(k)} \Omega^{[0k]}_{\mathrm{C}} = -\frac{\Sigma c}{2}
  |r^{\Hat{0}}| \sqrt{\frac{{\bf V}^{2}_{\mathrm{C}}}{[{\bf r}
  \times {\bf V}_{\mathrm{C}}]^{2}}} \; [{\bf r} \times {\bf
  V}_{\mathrm{C}}]_{(n)} \; .\eqno{(6.56)}
$$
The latter relation shows, that $s^{\mathrm{rot}}_{[0n]}$
represents a specific angular momentum of the point $\mathrm{M}$
relative to the center of inertia.

\parindent=24pt\ In so far as extended particle model in question
implies, that a particle has infinite extension in temporal axis,
expression (5.8) for proper angular momentum (spin) of the
particle should be redetermined in the form
$$
  S_{[\mu\nu]} = {\mathop{\int}\limits_{V}} \rho
  s^{\mathrm{rot}}_{[\mu\nu]} dV = {\mathop{\int}\limits_{D}}
  \rho^{(4)}(r^{\Hat{\alpha}}) s^{\mathrm{rot}}_{[\mu\nu]} d^{4}
  \mathrm{r} = {\mathop{\int}\limits_{V}} dV
  {\mathop{\int}\limits^{+\infty}_{-\infty}}
  \rho^{(4)}(r^{\Hat{\alpha}}) s^{\mathrm{rot}}_{[\mu\nu]}
  dr^{\Hat{0}} \; ,\eqno{(6.57)}
$$
where $\rho^{(4)}(r^{\Hat{\alpha}})$ is four-dimensional mass
density of the particle, $d^{4} \mathrm{r} = dr^{\Hat{0}} dV =
dr^{\Hat{0}} dr^{\Hat{1}} dr^{\Hat{2}} dr^{\Hat{3}}$ is an
elementary volume of four-dimensional domain $\it D$, occupied
with particle substance, which is bounded in this model by
hypersurfaces of light cone (6.49) and one-sheet hyperboloid
${\mathrm{r}}^{2} = h^{0}_{\Hat{\alpha}\Hat{\beta}}
r^{\Hat{\alpha}} r^{\Hat{\beta}} = 0$. One can see from (6.56),
that pseudovector ${\bf S} = \{ S_{[0n]} \}$ should be implied as
spin of the particle, interpreted as proper mechanical angular
momentum.

\parindent=24pt\ In our case transformations (6.9) do not exhaust
all transformations satisfying to condition (6.3). As an example
we can give a class of transformations, deriving from
five-dimensional interval (6.4). Analog of transformation (6.5)
may be written as~\cite{Tar1},~\cite{Tar3}
$$
  {\bf L} = {\bf l} + \frac{1}{R_{0}} \varphi (\alpha_{\omega})
  \sinh\sqrt{\alpha_{\omega}} \; \Hat{{\bf \rho}} +
  \frac{1}{R^{2}_{0}} \left ( \cosh\sqrt{\alpha_{\omega}} - 1
  \right) \varphi^{2} (\alpha_{\omega}) {\Hat{{\bf \rho}}}^{2} \;
  ,\eqno{(6.58)}
$$
where $\varphi (\alpha_{\omega})$ is arbitrary function on
$\alpha_{\omega}$, satisfying to condition $0 \leq \varphi^{-2}
(\alpha_{\omega}) < 1$, and
$$
  \varphi^{-2} (\alpha_{\omega}) = \frac{1}{2 R^{2}_{0}}
  \mathrm{Sp} {\Hat{{\bf \rho}}}^{2} =
  -\frac{g^{\Hat{5}\Hat{5}}}{R^{2}_{0}}
  g_{\Hat{\alpha}\Hat{\beta}} \rho^{\Hat{\alpha}}
  \rho^{\Hat{\beta}} = -\frac{1}{R^{2}_{0}}
  h^{0}_{\Hat{\alpha}\Hat{\beta}} \rho^{\Hat{\alpha}}
  \rho^{\Hat{\beta}} = -\frac{{\mathrm{\rho}}^{2}}{R^{2}_{0}} =
  \beta^{2} \; .\eqno{(6.59)}
$$
It should be express $\alpha_{\omega}$ through $\beta^{2}$ from
(6.59) and substitute it to (6.58). Then transformation law of
vectors $r^{\Hat{\alpha}}$ would be derived similar to formula
(6.9).

\parindent=24pt\ Moreover, using of transformation (6.58) admits a
generalization of condition (6.3) in the form
$$
  -R^{2}_{k+1} < \mathrm{r}^{2} = h^{0}_{\Hat{\alpha}\Hat{\beta}}
  r^{\Hat{\alpha}} r^{\Hat{\beta}} < -R^{2}_{k} \; , \; \; k = 0,
  1, 2, ..., K \; ,\eqno{(6.60)}
$$
which leads to layered structure of internal space~\cite{Tar4}. It
is quite obvious also, that to construct a real metric one can use
nonlinear transformations, based on principles, which differ from
the principle, used here.

\section{ Conclusive notes}

\parindent=24pt\ \; \; \; Essential moment in our attempt to
describe an extended elementary particle is using a postulate of
geodesic lines~\cite{DeSit1}-\cite{DeSit3}. Applying it to an
extended particle we specified a concept of the center of inertia
by condition that its motion should pass along geodesic line of
external space. Geodesic lines of internal space should represent
a motion of substance inside of a particle relative to the center
of inertia. Physically it is expressed in independency of internal
motions on external motion. Geometrically it turns out to be
expressed in terms of degenerated embedding spaces. If it is
hardly valid for macroscopic extended bodies, for internal and
external motions are very strongly coupled. But unobservability of
internal motions and constancy of some characteristics of
particles at least at not very large energies gives possibility to
postulate this assumption just for elementary particles. This is a
basic distinction of suggested method from the description,
adopted in General Relativity, equations of which are used only
for interpretation of the metric.

\parindent=24pt\ Here we considered only a one-particle problem.
Therefore, it was not necessary to introduce such characteristic
of the particle as mass and charge. Apparently, they must appear
as parameters of interaction between two extended objects. But, in
return, there appears a quantity $R_{0}$, characterizing a size of
a particle. Instead of quantum-mechanical spin we consider
mechanical proper angular momentum, whose presence or absence of
association with spin should be determined later about. We think
such association to be existing, and this method may be applied to
stable particles, such as electron and proton, but when their
interactions are considering it should be modified.


\begin{thebibliography}{99}

\bibitem[1]{Dis} Discussion on rigid body. \; In: Einshteinovskiy
sbornik, \; 1975-1976. -- M.: Nauka, 1978. -- pp. 287-350 (in
Russian).

\bibitem[2]{Fre} Frenkel Ya.I. \; Elektrodinamika. \; Sobr. \; izbr.
\; tr., \; t. 1. -- M.-L.: AN SSSR, 1956 (in Russian).

\bibitem[3]{Izma} Izmailov S.V. On quantum relativistic theory of
particles, possessing internal rotational degrees of freedom. \;
I. -- JETP, 1947, {\bf 17}, vyp. 7, pp. 629-647 (in Russian).

\bibitem[4]{Tak1} Takabayashi T. Relativistic Particle with
Internal Rotational Structure. -- Nuovo Cim., 1959, {\bf 13}, no.
3, pp. 532-554.

\bibitem[5]{Tak2} Takabayashi T. Theory of Relativistic Rotators
and Elementary Particles. -- Progr. Theor. Phys., 1960, {\bf 23},
no. 5, pp. 915-941.

\bibitem[6]{Tak3} Takabayashi T. Internal Degrees of Freedom and
Elementary Particles. -- Progr. Theor. Phys., 1961, {\bf 25}, no.
6, pp. 901-938.

\bibitem[7]{Wey} Weyssenhoff J., Raabe A. Relativistic Dynamics of
a Spin-fluids and Spin-particles. -- Acta Phys. Polon., 1947, {\bf
9}, no. 1, pp. 7-18.

\bibitem[8]{Dix1} Dixon W.G. On a Classical Theory of Charged
Particles with Spin andthe Classical Limit of the Dirac Equation.
-- Nuovo Cim., 1965, {\bf 38}, no. 4, pp. 1616-1643.

\bibitem[9]{Mar} Markov M.A. On "four-dimensionally extended"
electron in relativistic quantum domain. -- JETP, 1940, {\bf 10},
vyp. 12, pp.1311-1338 (in Russian).

\bibitem[10]{Yuk} Yukawa H. Space-Time Description of Elementary
Particles. -- Proc. \; Int. \; Conf. \; Elem. \; Part., 1965,
Kyoto, pp. 139-158. --– This report results earlier investigations
on bilocal theory.

\bibitem[11]{Tak4} Takabayashi T., Kojima S. Relativistic
Mechanics of Interacting Particles and Multi-Local Theory. I. The
Bilocal Case. -- Progr. Theor. Phys., 1977, {\bf 57}, no. 6, pp.
2127-2143; Takabayashi T., idem, II. Trilocal and More General
Cases. -- Progr. Theor. Phys., 1977, {\bf 58}, no. 4, pp.
1299-1315.

\bibitem[12]{Nak} Nakano T. A Relativistic Field Theory of an
Extended Particle. -- Progr. Theor. Phys., 1956, {\bf 15}, no. 4,
pp. 333-368.

\bibitem[13]{Hara} Hara O., Goto T. Extended Particle Model of
Elementary Particles. -- Progr. Theor. Phys., 1968, Suppl. no. 41,
pp. 56-129.

\bibitem[14]{Grot} Grot R.A., Eringen A.C. Relativistic Continuum
Mechanics, Part I. -- Int. J. Engng Sci., 1966, {\bf 4}, no. 6,
pp. 611-638; Part II, ibid., pp. 639-669.

\bibitem[15]{Tul1} Tulczyjew W. Equation of Motion of Rotating
Bodies in General Relativity Theory. -- Acta Phys. Pol., 1959,
{\bf 18}, no. 1, pp. 37-55.

\bibitem[16]{Tul2} Tulczyjew W. Motion of Multipole Particles in
General Relativity Theory. -- Acta Phys. Pol., 1959, {\bf 18}, pp.
393-409.

\bibitem[17]{Dix2} Dixon W.G. Extended Bodies in General
Relativity: Their Description and Motion. -- Proc. Int. School
Phys. E.Fermi, Course LXVII, 1979, pp. 156-219.

\bibitem[18]{Ryab} Ryabushko A.P. Motion of Bodies in General
Relativity Theory. -- Minsk: Vysheishaya shkola, 1979. -- 240 pp.
(in Russian).

\bibitem[19]{Math} Mathisson M. Neue Mechanik materieller Systeme.
-- Acta Phys. Pol., 1937, {\bf 6}, no. 3, pp. 163-200.

\bibitem[20]{Pap} Papapetrou A. Spinning Test-Particle in General
Relativity. I. -- Proc. Roy. Soc., 1951, {\bf 209A}, no. 1097, pp.
248-258; Corinaldesi E., Papapetrou A. idem, II, ibid., pp.
259-268.

\bibitem[21]{Kam} Kamimura K. Differential Form Description of
Geometrical Model of Hadrons in Space-Time. -- Progr. Theor.
Phys., 1981, {\bf 65}, no. 5, pp. 1704-1718.

\bibitem[22]{Pry} Pryce M.H.L. The Mass-centre in the Restricted
Theory of Relativity and its Connexion with the Quantum Theory of
Elementary Particles. -- Proc. Roy. Soc., 1948, {\bf 195A},
pp.62-81.

\bibitem[23]{DeSit1} De Sitter W. On Einstein's Theory of
Gravitation, and its Astronomical Consequences. First paper. --
Mon. Not. Roy. Astr. Soc., 1916, {\bf 76}, no. 9, pp. 699-728.

\bibitem[24]{DeSit2} De Sitter W. On Einstein's Theory of
Gravitation, and its Astronomical Consequences. Second paper. --
Mon. Not. Roy. Astr. Soc., 1916, {\bf 77}, no. 2, pp. 155-184.

\bibitem[25]{DeSit3} De Sitter W. On Einstein's Theory of
Gravitation, and its Astronomical Consequences. Third paper. --
Mon. Not. Roy. Astr. Soc., 1917, {\bf 78}, no. 1, pp. 3-28.

\bibitem[26]{Pet} Petrov A.Z. On simulating of paths of test
bodies in the field of gravitation. -- Dokl. AN SSSR, 1969, {\bf
186}, no. 6, pp. 1302-1305 (in Russian).

\bibitem[27]{Lan} Landau L.D., Lifshits E.M. Mechanics. -- L.:
Pergamon Press, 1980.

\bibitem[28]{Halb1} Halbwachs F., Hillion P., Vigier J.-P.
Quadratic Lagrangians in Relativistic Hydrodynamics. -- Nuovo
Cim., 1959, {\bf 11}, no. 6, pp. 882-883.

\bibitem[29]{Halb2} Halbwachs F., Hillion P., Vigier J.-P.
Internal Motions of Relativistic Fluid Masses. -- Nuovo Cim.,
1960, {\bf 15}, no. 2, pp. 209-232.

\bibitem[30]{Sed} Sedov L.I. Mechanics of continuous media. Vol.
I. -- M.: Nauka, 1976. -- 535 pp.

\bibitem[31]{Scho} Schouten J.A. Tensor Analysis for Physicists.
-- Oxford: The Clarendon Press, 1951.

\bibitem[32]{Szek} Szekeres P. Embedding Properties of General
Relativistic Manifolds. -- Nuovo Cim., 1966, {\bf 43A}, no. 4, pp.
1062-1076.

\bibitem[33]{Jan} Jankiewicz Cz. On degenerating Riemannian
spaces. -- Bull. Acad. Sci. Pol., Section III, 1954, {\bf 2}, no.
7, pp. 305-308 (in Russian).

\bibitem[34]{Born} Born M., Fuchs K. Reciprocity. II. Scalar Wave
Functions. -- Proc. Roy. Soc. Edinb., 1939-1940, {\bf A60}, pp.
100-116.

\bibitem[35]{Tar1} Tarakanov A.N. On Real and Complex "Boost"
Transformations in Arbitrary Pseudo-Euclidean Spaces. -- Teor.
Mat. Fiz. 1976, {\bf 28}, no. 3, pp. 352-358 (in Russian); Theor.
\& Math. Phys. (USA), 1978, {\bf 18}, p. 838 (in English).

\bibitem[36]{Tar2} Tarakanov A.N. Homogeneous Spacetimes as Models
for Isolated Extended Objects. --  Proceedings of 5th
International Conference Bolyai-Gauss-Lobachevsky: Non-Euclidean
Geometry In Modern Physics (BGL-5), 10-13 Oct 2006, Minsk,
Belarus. -- pp.304-314. -- http://www.arxiv.org: hep-th/0611149 --
12 pp.

\bibitem[37]{Bog1} Bogush A.A., Fyodorov F.I. Generalized
Kronecker symbols. -- Dokl. AN BSSR, 1968, {\bf 12}, no. 1, pp.
21-24 (in Russian).

\bibitem[38]{Bog2} Bogush A.A., Moroz L.G. Introduction to the
Theory of Classic Fields. -- Minsk: Nauka i tehnika, 1968. -- 388
pp. (in Russian).

\bibitem[39]{Kra} Kramer D., Stephani H., MacCallum M., Herlt E.
Exact Solutions of Einstein's Field Equations. -- Berlin:
Deutscher Verlag der Wissenschaften, 1980.

\bibitem[40]{Deh1} Dehmelt H. Single Atomic Particle Forever
Floating at Rest in Free Space: New Value for Electron Radius. --
Physica Scripta, 1988, {\bf T22}, pp. 102-110.

\bibitem[41]{Deh2} Dehmelt H. Experiments with an
isolated subatomic particle in rest. Nobel Lecture, December 8,
1989. -- Physica Scripta, 1991, {\bf T34}, pp. 47-51.

\bibitem[42]{Tar3} Tarakanov A.N. On parametrization of boosts in
multidimensional pseudo-Euclidean spaces. In: Covariant methods in
theoretical physics. Physics of elementary particles and
relativity theory, vyp. 5. -- Minsk: Institute of physics of NANB,
2001. -- pp. 137-142 (in Russian).

\bibitem[43]{Tar4} Tarakanov A.N. Generalized boost
transformations with shell structure of parameter space. In:
Gravitation and electromagnetism. -- Minsk: Izd.
"Universitetskoye", 1987, pp. 149-156 (in Russian).


\end{thebibliography}
\end{document}